\documentclass[12pt]{article}

\usepackage{epsfig,amsmath,amssymb,latexsym,axodraw}

\providecommand{\beqa}{\begin{eqnarray}}
\providecommand{\eeqa}{\end{eqnarray}}
\providecommand{\om}{{\omega}}
\providecommand{\Cbar}{\overline{C}}
\providecommand{\Ibar}{\overline{I}}
\providecommand{\Jbar}{\overline{J}}
\providecommand{\Kbar}{\overline{K}}

\providecommand{\Mbar}{\overline{M}}

\providecommand{\Sbar}{\overline{S}}
\providecommand{\Tbar}{\overline{T}}

\providecommand{\Wbar}{\overline{W}}
\providecommand{\ibar}{\bar{\imath}}

\providecommand{\hbarr}{\bar{h}}

\providecommand{\Lambdabar}{\bar{\Lambda}}
\providecommand{\Ktil}{\tilde{K}}

\providecommand{\we}{\wedge}
\providecommand{\tr}{\text{tr}}

\def\cG{{\cal{G}}}
\def\cH{{\cal{H}}}
\def\cL{{\cal{L}}}
\def\cN{{\cal{N}}}
\def\cO{{\cal{O}}}
\def\cS{{\cal{S}}}
\def\cV{{\cal{V}}}

\def\Mtil{{\tilde{M}}}
\def\mtil{{\tilde{m}}}
\def\Re{{{\text{Re}}}}
\def\Im{{{\text{Im}}}}

\numberwithin{equation}{section}

\begin{document}

\thispagestyle{empty}
\rightline{UMD-PP-04-016, HU-EP-04/14}

\begin{center}
{\bf \LARGE De Sitter Vacua from Heterotic M-Theory}
\end{center}

\vspace{1.3truecm} \centerline{Melanie Becker
$^{a,}$\footnote{melanieb@physics.umd.edu}, Gottfried Curio
$^{b,}$\footnote{curio@physik.hu-berlin.de}, Axel Krause
$^{a,}$\footnote{krause@physics.umd.edu}}
\vspace{.6truecm}

{\em \centerline{$^a$ Department of Physics, University of
Maryland,} \centerline{College Park, MD 20742, USA}}
\vspace{.3truecm} {\em \centerline{$^b$ Humboldt-Universit\"at zu
Berlin,} \centerline{Institut f\"ur Physik, D-12489 Berlin,
Germany}}

\vspace{1.0truecm}

\begin{abstract}
It is shown how metastable de Sitter vacua might arise from
heterotic M-theory. The balancing of its two non-perturbative
effects, open membrane instantons against gaugino condensation on
the hidden boundary, which act with opposing forces on the
interval length, is used to stabilize the orbifold modulus
(dilaton) and other moduli. The non-perturbative effects break
supersymmetry spontaneously through F-terms which leads to a
positive vacuum energy density. In contrast to the situation for
the weakly coupled heterotic string, the charged scalar matter
fields receive non-vanishing vacuum expectation values and
therefore masses in a phenomenologically relevant regime. It is
important that in order to obtain these de Sitter vacua we are not
relying on exotic effects or fine-tuning of parameters. Vacua with
more realistic supersymmetry breaking scales and gravitino masses
are obtained by breaking the hidden $E_8$ gauge group down to
groups of smaller rank. Also small values for the open membrane
instanton Pfaffian are favored in this respect. Finally we outline
how the incorporation of additional flux superpotentials can be
used to stabilize the remaining moduli.
\end{abstract}

\noindent
PACS: 04.65.+e, 11.25.Mj, 11.25.Yb\\
Keywords: De Sitter Vacua, Heterotic M-Theory\\
hep-th/0403027

\newpage
\pagenumbering{arabic}

\section{Introduction and Conclusions}

In view of current astronomical data \cite{WMAP} which are in nice
agreement with a dark energy component generated by a cosmological
constant, modern theoretical physics faces the challenge of
finding realistic non-supersymmetric vacua with positive vacuum
energy. Indeed these seem to be the right class of vacua not only
today but also during an early epoch of inflation with however
vastly different vacuum energies. The search for de Sitter vacua
in ordinary supergravity theories dates back to some rather early
papers (see e.g. \cite{GZ} and references therein). More recently
the connection between supergravity theories and de Sitter vacua
was discussed in \cite{FTP}, \cite{BS}, \cite{GdS} and other
interesting ideas on how to obtain de Sitter vacua from
string-theory appeared in \cite{KKLT}, \cite{dSLit}, \cite{SS}.

Progress towards a straightforward derivation of 4d de Sitter
spaces or more generally 4d accelerated cosmologies from a
standard reduction of 10d effective string-theories was hampered
by the No-Go theorems presented in \cite{NK} and \cite{NoGo}.
However, ways around these theorems were found -- either in the
way of incorporating instantons \cite{CK2}, \cite{KKLT} or
time-dependent hyperbolic compact internal spaces \cite{TW} (which
arise as special S-brane \cite{GS} solutions with vanishing flux
\cite{NO}; S-branes were connected to accelerated cosmologies
e.g.~in \cite{SBC}). While the former case leads to 4d de Sitter
vacua the latter situation leads to accelerating 4d cosmologies
with an equation of state $p=w\rho$, where $w=(4-n)/(3n)$ with $n$
the number of internal compact dimensions, $p$ being pressure and
$\rho$ the density. Unfortunately, the tight observational
constraint of $w<-0.78$ \cite{WMAP} resulting from the combined
WMAP (Wilkinson Microwave Anisotropy Probe) data, supernova
observations, HST (Hubble Space Telescope) data and 2dFGRS (2dF
Galaxy Redshift Survey) large scale structure data evaluated under
the premise that $w > -1$ cannot be satisfied by any
$n\in\mathbf{N}$. Moreover without this premise a value of
$w=-0.98\pm 0.12$ is favored \cite{WMAP} by the combined data
which clearly points towards a cosmological constant which gives
$w=-1$ (which again cannot be obtained for any positive integer
$n$). We are therefore motivated to address in this paper the
question of how 4d de Sitter spaces can arise from string/M-theory
leaving the question of a realistic cosmology with these spaces
for future work.\footnote{For a recent interesting proposal to
realize inflation in M-theory see e.g. \cite{DHHK}.}

The interesting question on how to embed de Sitter solutions into
string theory has been lately addressed in the literature in the
context of string theory and M-theory compactifications with
non-vanishing fluxes. In fact, de Sitter spaces in IIB
string/F-theory in the presence of NSNS and RR fluxes were found
recently in the work of \cite{KKLT} and \cite{SS} building on
results of \cite{GKP}. In order to stabilize the moduli fields the
work of \cite{KKLT} used NSNS and RR fluxes while anti-D3-branes
were introduced to achieve the necessary supersymmetry breaking.

From the phenomenological point of view (see
e.g.~\cite{WWarp},\cite{HetMPheno}) it is, of course, more
interesting to consider heterotic M-theory instead of Type IIB
theory. It has been known for quite a while that de Sitter vacua
can occur in Calabi-Yau (CY) flux compactifications of heterotic
M-theory when non-perturbative effects are included \cite{CK2}.
Our goal in this paper is to establish the de Sitter vacua found
in \cite{CK2} more rigorously by using instead of the linearized
background of \cite{WWarp} the full non-linear background of
\cite{CK1},\cite{CK3} and moreover by including charged matter
fields, which is of interest for particle phenomenology as well as
cosmology. Recall that in heterotic M-theory the stabilization of
the orbifold length and the corresponding axion requires a
four-form $G$-flux component $G_{(2,2,0)}$ of M-theory which is
projected out in the weakly coupled string-theory limit. Thus the
stabilization mechanism of \cite{CK2} is a truly strongly coupled
one. This mechanism is in a sense complemental to the moduli
stabilization mechanism of the weakly coupled heterotic string
that has been recently proposed in the literature \cite{BD},
\cite{TT} and \cite{CL} based on a non-trivial
$G_{(1,2,1)}=H_{1,2}$ component. We will comment towards the end
of the paper on the relation between our paper and the results
obtained in the previous references in the context of the weakly
coupled heterotic string. In fact, we shall see that the
stabilization of the radial modulus obtained in \cite{BD} expands
very nicely the stabilization of the moduli fields achieved in
heterotic M-theory and we will borrow the results from \cite{BD}
in order to fix most (and maybe all) of the moduli fields. Another
way to stabilize at least some of the remaining moduli is to
consider alternatively a balancing of the $H_{0,3}$ component of
the Neveu-Schwarz $H$-flux against gaugino condensation as was
done for the weakly coupled heterotic string in \cite{GKLA}. In
fact, this might be an easier starting point to incorporate
further moduli, as the internal CY manifold still remains
K\"ahler. Work in this direction is in progress.

Getting back to the strongly coupled theory, we shall see that
non-perturbative open membrane instanton effects contributing to
the superpotential break supersymmetry spontaneously, leading to a
positive scalar potential. Therefore, our primary goal here will
be to establish 4d de Sitter vacua in heterotic M-theory in the
presence of charged matter fields $C$ and to explore the global
potential landscape beyond what was possible in \cite{CK2} due to
the limitations of the linearized background to show that (in
agreement with general expectations \cite{dS1}, \cite{dS2},
\cite{dS3}, \cite{KKLT}) the arising de Sitter minima are actually
false metastable vacua. The possibility to study the global
potential profile arose only recently after it was understood
\cite{CK3} (based on \cite{CK1}) how to extend the linearized
background for flux compactifications of heterotic M-theory to the
full non-linear background and still preserve supersymmetry. Only
with this extension will it be possible to study the potential for
arbitrary large orbifold-modulus $\cL$ and thus to establish the
expected runaway behaviour in the decompactification limit (see
also the discussion in \cite{KSUSY03}). The linearized background
has been used recently in \cite{BO} to find supersymmetric Anti de
Sitter vacua in heterotic M-theory.

Let us summarize briefly some of the results obtained in this
paper. In order to find 4d de Sitter vacua in our context the
orbifold length and the volume modulus have to be stabilized. We
will find that one can naturally, i.e. without invoking
hierarchically large or small values for some parameters stabilize
the orbifold modulus $\cL_0$ close to the maximally allowed value
$\cL_{max}$, which is quite satisfying as this position leads to
the correct 4d Newton's Constant once the GUT-scale and the GUT
gauge coupling attain their usual values.

The stabilization of the orbifold length modulus is exemplified in
figure \ref{OMGCPotential} appearing later on in the paper. The
logarithmic plot of figure \ref{OMGCPotential}, displaying the
full behavior of the potential in the whole interval
$\cL\in[0,\cL_{max}]$ (whose bound arises from a true physical
singularity not from an artifact of the linear approximation),
shows nicely that the minimum is caused by a balancing between two
distinct non-perturbative effects possessing different monotony
behavior: open membrane instantons and gaugino condensation. The
precise equation for this balancing will be derived in section 4,
but we will show there also that it is well approximated by the
simplified condition
\beqa
 e^{-\Re T} = e^{-\frac{1}{C_H}\Re S+\frac{\gamma}{C_H}\Re T}
\eeqa
which expresses clearly the {\em balance between open membrane
instantons $(\sim e^{-\Re T})$ and gaugino condensation} $(\sim
e^{-\frac{1}{C_H}\Re S+\frac{\gamma}{C_H}\Re T})$. Moreover one
finds that at leading order in $1/\cV$ and $1/\cV_{OM}$ e.g.
\beqa
 D_S W = -\frac{W_{GC}}{C_H} \ne 0 \; ,
\eeqa
which shows that at the location of the minimum in particular
$D_SW$ is non-vanishing, i.e. {\em supersymmetry is broken
spontaneously} through F-term expectation values. Moreover we find
that the resulting supersymmetry breaking scale and gravitino mass
can be brought close to the phenomenologically relevant regime
when the hidden $E_8$ is broken down to smaller gauge groups.
Moreover small values for the Pfaffian $h$ arising in the open
membrane superpotential are favored in this respect.

Until now for this class of vacua not all moduli were stabilized
explicitly thus the de Sitter vacua could potentially be unstable
in some directions. Nevertheless the vacua would still have
positive vacuum energy. The stabilization of these remaining
moduli might however be achieved by incorporating in addition
$H$-flux superpotentials which we will outline in the last part of
the paper. We will concentrate on the simplest case possible,
namely on the standard vacua without additional M5 branes but
including the most dominant non-perturbative effect coming from
open membrane instantons stretching between the two boundaries. We
will treat the case with an additional M5 brane elsewhere. We will
show that when additional non-perturbative effects coming from
gaugino condensation on the hidden boundary are taken into account
it is possible to stabilize in addition the charged matter fields.
This is a rather promising result for particle phenomenology as
for a long time one of the major drawbacks of heterotic string
theory was the presence of massless charged matter fields (see
e.g. \cite{Nano}and \cite{BinGai}). We will see that the vacuum
expectation values (vev) for these scalars lie in a
phenomenologically interesting range.

The organization of the paper is as follows. In section 2 we
compute the effective potential resulting from compactifications
of heterotic M-theory on a manifold $X\times S_1/\mathbf{Z}_2$,
where $X$ is the internal CY three-fold taking non-trivial fluxes,
open menmbrane instantons and gaugino condensation into account.
In section 3 we analyze the scalar potential obtained in the
previous section without taking gaugino condensation into account
and show that while the charged scalars obtain a non-trivial vev
the orbifold length does not get stabilized. In section 4 we
include the effects of gaugino condensation and show that the
axions and both the orbifold length and the charged matter field
obtain a non-trivial vev. We finish in section 5 with a discussion
on the connection to the moduli stabilization mechanisms for the
weakly coupled heterotic string recently proposed in the
literature \cite{BD}, \cite{TT}, \cite{CL} and \cite{GKLA}. In
fact we borrow some results of these papers e.g. the stabilization
of the radial modulus achieved in \cite{BD} to fix many (and maybe
all) of the moduli fields appearing in our compactifications.
Clearly a more direct analysis has to be done to clarify which
moduli fields are precisely stabilized. This question shall not be
addressed in this paper and will be left for future work.

\section{The Effective 4D Heterotic M-Theory}

Our starting point are compactifications of heterotic M-theory
\cite{HW1}, \cite{HW2} on an internal seven-space
\beqa
X\times \mathbf{S}^1/\mathbf{Z}_2
\eeqa
where $X$ is a Calabi-Yau (CY) threefold. The resulting effective
4d theory is described by an N=1 supergravity \cite{LOW1} which is
completely determined by the gauge kinetic function, the K\"ahler-
and the superpotential for the occuring moduli. As long as one
switches on only the $G_{(2,2,0)}$ (the first number counts
holomorphic, the second antiholomorphic CY indices and the third
the orbifold index) component of the 11d supergravity four-form
field-strength the resulting 7d flux compactification background
is given by a warped geometry whose 6d piece is a conformal
deformation of $X$ \cite{WWarp}, \cite{CK1}, \cite{CK3}. One
therefore has $h^{(1,1)}$ complex moduli $T^i$ which correspond to
deformations of the K\"ahler class $\omega$ of $X$ and $h^{(1,2)}$
complex moduli $Z^\alpha$ describing the deformations of the
complex structure of $X$.

Notice, however, that when switching on a $G_{(1,2,1)}$ component
which corresponds to a Neveu-Schwarz background $H_{NS}$ in the
10d limit where the orbifold length shrinks to zero, it is known
that then $X$ becomes a non-K\"ahler manifold which is no longer
conformal to a CY \cite{NK} and whose moduli are not explicitly
known. We will therefore in the main part assume that
$G_{(1,2,1)}=H_{1,2}=0$ and will comment towards the end on the
consequences of including it. The fact that in heterotic M-theory
one can set $G_{(1,2,1)}$ to zero therefore allows us to avoid the
complications which arise in the weakly coupled heterotic string
but nevertheless study the implications of the $G_{(2,2,0)}$
component for the stabilization of the universal moduli. The
situation is therefore similar to the type IIB case, where a
nontrivial $H_{NS}$ also merely leads to a conformal deformation
of the Calabi-Yau.

In addition to the moduli described above one has the volume
modulus $S$ and the charged matter $C^I$. Here $I$ represents a
multi-index $({\cal R},i)$ running over the representations ${\cal
R}$ of the unbroken visible gauge group $G$ (where $G\times
G_{hol}\subset E_8$ and $G_{hol}$ includes the gauge instanton's
holonomy), $i=1,\hdots,\text{dim}\,H^1(X,V_{\cS})$ (respectively
also over $n=1,\hdots,\text{dim} \, {\cal R}$ when we are
referring to the charged scalar components inside the
representation; they are not to be confused with the $h^{(1,1)}$
neutral scalars given by the K\"ahler moduli) where the $V_{\cS}$
are those vector bundles into which the gauge bundle decomposes
when the {\bf 248} of $E_8$ is decomposed under $G\times G_{hol}$.
In this paper we will take a CY compactification with
$h^{(1,1)}=1$ for simplicity.

The moduli are then combined into the following 4d, N=1 chiral
superfields
\begin{alignat}{3}
&S = \cV(\cL)+i\sigma_S \\
&T = \cV_{OM}(\cL)+i\sigma_T \\
&C^I, \quad Z^\alpha
\end{alignat}
Here $\cV_{OM}$ describes the normalized volume of an open
membrane instanton stretching between both boundaries and wrapping
a holomorphic curve $\Sigma$ inside the CY (for simplicity we will
assume that the instanton wraps $\Sigma$ only once)
\beqa
\label{VOM}
\cV_{OM}(\cL) = \cL \left(\frac{6\cV(\cL)}{d}\right)^{1/3} \; ,
\eeqa
where $d$ is the CY intersection number. The dimensionless moduli
$\cV(\cL)$, $\cL$ are related to the dimensionful CY volume
$V(x^{11})$ and the orbifold length $L$ through ($x^{11}\in [0,L]$
is the 11d orbifold coordinate)
\beqa
\label{VLMod}
\cV(\cL) = \frac{\langle V(x^{11}) \rangle}{v}
= \frac{1}{vL}\int_0^L dx^{11} V(x^{11})  \; , \qquad
\cL = \frac{L}{l}
\eeqa
where $v$ and $l$ are two conveniently chosen dimensionful
reference values \cite{MPS}
\beqa
\label{Units}
v = 8\pi^5l_{11}^6 \; , \qquad l = 2\pi^{1/3}l_{11}
\eeqa
given in terms of the 11d Planck-length $l_{11}$ which itself is
related to the 11d gravitational coupling constant $\kappa$
through
\beqa
\label{Planck}
2\kappa^2 = (2\pi)^8l_{11}^9 \; .
\eeqa
The axions $\sigma_S$ and $\sigma_T$ arise from two different
components of the 11d three-form potential $C_{AB11}$. While
$\sigma_S$ comes from the ususal 4d dualization of $C_{\mu\nu
11}$, one obtains $\sigma_T$ as the coefficient from the expansion
of $C_{lm11}$ in terms of the single base element of $H^{1,1}(X)$
(remember that $h^{(1,1)}=1$).

\subsection{The K\"ahler-Potential}

We won't need the gauge kinetic functions in the following. So let
us specify first the K\"ahler-potential for the above moduli. For
the $S$ and $T$ moduli the K\"ahler-potential reads \cite{LOW1}
\beqa
\label{KST}
K_{(S)} = -\ln(S+\Sbar) \; , \qquad K_{(T)} =
-\ln\Big(\frac{d}{6}(T+\Tbar)^3\Big)
\eeqa
while for the $C$'s it is given by
\beqa
\label{KC}
K_{(C)} = \left( \frac{3}{T+\Tbar} + \frac{2\beta_v}{S+\Sbar}
\right) H_{I\Jbar}C^I {\overline{C}}^{\Jbar}
\eeqa
at leading order in the $C^I$. The positive-definite metric
$H_{I\Jbar}$ depends only on the complex structure and bundle
moduli \cite{MPS,LOW4} while the instanton number
$\beta_v\in\mathbf{Z}$ of the visible boundary is given by the
expansion coefficient of the visible boundary second Chern-classes
\beqa
\label{Chern}
c_2(F_v)-\frac{1}{2}c_2(R)
= \frac{-\tr F_v\we F_v + \frac{1}{2}\tr R\we R}{8\pi^2}
= \beta_v[\Sigma_1] \; .
\eeqa
Because $h^{(1,1)}=1$, there is only one basis element $\Sigma_1$
of the second homology group $H_2(X,\mathbf{Z})$ whose Poincar\'e
dual four-form is $[\Sigma_1]$. An analogous expansion involving
the hidden boundary gauge field defines the hidden sector
instanton number $\beta_h\in\mathbf{Z}$. Anomaly cancelation
demands that
\beqa
\beta_v+\beta_h = 0
\eeqa
which means that either $\beta_v$ or $\beta_h$ has to be negative.
In the phenomenologically relevant case where the volume of the
deformed CY decreases from the visible towards the hidden
boundary, $\beta_v$ is positive \cite{WWarp}
\beqa
\beta_v > 0
\eeqa
which is the case we will consider henceforth.

Besides these contributions to the K\"ahler-potential, there is
also the contribution from the complex structure moduli
\beqa
K_{(Z)} = -\ln \Big(-i\int_X\Omega\we\overline{\Omega}\Big)
\eeqa
and the K\"ahler-potential $K_{(A)}$ for the vector bundle moduli.
We will consider the complex structure moduli as `frozen' in this
paper and address their stabilization in a separate publication.
The contribution from the vector bundle moduli, $K_{(A)}$, is
considerably suppressed. In \cite{BO} it was shown to be
generically smaller by a factor $10^{-5}$ as compared to
$K_{(S)},K_{(T)}$.

The K\"ahler-potential described so far, in particular the
contributions (\ref{KST}),(\ref{KC}) which we will employ later
on, comprises those terms which are universally present in all
heterotic compactifications. As such they would also occur in
compactifications over the background derived in \cite{CK1},
\cite{CK3} (see also \cite{K1}) which generalizes and extends the
flux compactification background of \cite{WWarp} in such a way
that the background geometry becomes trustworthy until a naked
singularity is hit at some finite critical value of the orbifold
coordinate
\beqa
x^{11}_0 = \frac{l}{\cG_v}
\eeqa
determined by the dimensionless visible boundary flux-parameter
$\cG_v$ which will be defined below in (\ref{GFlux1}). At
$x^{11}=x_0^{11}$ the `classical' CY volume vanishes. The
derivation of the effective 4d heterotic M-theory action as
presented in the literature \cite{LOW1} took as a starting point
the flux compactification background of \cite{WWarp}. This
background was obtained as a solution to the 11d gravitino
Killing-spinor equation under the assumption that the warp-factor
of the background geometry stays small and could be used as a
dimensionless expansion parameter. Consequently the solution is a
perturbative one which is linear in the warp-factor and neglects
all higher powers in the warp-factor. The imposition of the
smallness of the parameter
\beqa
\label{Eps}
\epsilon = \frac{2\pi L}{3V_v^{2/3}}
\Big(\frac{\kappa}{4\pi}\Big)^{\frac{2}{3}}
= \frac{2\cL}{3\cV_v^{2/3}} = \frac{2}{3}
\left(\frac{d}{6}\right)^{1/3} \frac{\cV_{OM}}{(\cV_v^2\cV)^{1/3}}
\simeq \frac{\cV_{OM}}{\cV}\left(\frac{d}{6}\right)^{1/3}
\eeqa
on the effective 4d theory, where
\beqa
\cV_v = \frac{V_v}{v}
\eeqa
is the dimensionless, normalized CY volume $V_v$ on the visible
boundary, stems directly from this linearized background solution.
Namely to ensure that the warp-factor stays small one has to
guarantee that its absolute value stays small which is
upper-bounded by $\epsilon$. Therefore, though phenomenology
requires an $\epsilon = \cO(1)$ \cite{BD1}, the perturbative
background demands a perturbatively small $\epsilon$.

It is however known how to extend the linear background solution
to incorporate the required higher order (in the warp-factor)
correction terms demanded by the full non-linear 11d gravitino
Killing-spinor equation, hence by supersymmetry \cite{CK1,CK3}. As
this extended full background leads to a manifest positive
Riemannian metric and positive CY volume over the full moduli
space (which is not the case for the linearized background), it is
the better starting point for an analysis of the effective
potential. This even more so because the extended background can
be used reliably until one hits the singularity at $x_0^{11}$ and
therefore stabilization of the orbifold modulus $\cL$ in the
phenomenologically favoured regime, where $\epsilon = \cO(1)$,
becomes feasible.

Though the complete reduction of heterotic M-theory over the
extended full background to obtain the corresponding 4d effective
theory hasn't been carried out yet \cite{ConstK}, it is
nevertheless clear that the universal structure of the
K\"ahler-potential (\ref{KST}),(\ref{KC}) or standard composition
of moduli into $N=1$, 4d chiral superfields won't change. What
will change, is the functional dependence of $\cV(\cL)$ on $\cL$
which becomes quadratic instead of linear as we will see below.
The extended full background of \cite{CK1},\cite{CK3} with its
manifestly positive metric and CY volume will be used in the
subsequent derivation of de Sitter vacua.

\subsection{The Superpotential}

The second ingredient needed to determine the potential is the
superpotential which consists of a perturbative and a
non-perturbative piece
\beqa
\label{W1}
W = W_{tree}+W_{non-pert} \; .
\eeqa
The perturbative piece is given by the standard cubic
superpotential \cite{LOSW}
\beqa
\label{WT}
W_{tree}
= \Lambda_{IJK} C^I C^J C^K
= \frac{4\pi\sqrt{2}}{3}\lambda_{IJK} C^I C^J C^K
\eeqa
where $\lambda_{IJK}$ are the Yukawa couplings. In general the
Yukawa couplings are quasi-topological, i.e.~they depend on the
complex structure and bundle moduli only \cite{LOW4}.

Leaving aside M5 branes in this paper, the non-perturbative
contribution to the superpotential comes from open membrane
instantons which stretch between the two boundaries and gaugino
condensation on the hidden boundary
\beqa
\label{WNP}
W_{non-pert} = W_{OM}+W_{GC} \; .
\eeqa
The former is described by a superpotential \cite{MPS,LOPR}
\beqa
\label{WOM}
W_{OM}=h e^{-T}
\eeqa
where the Pfaffian $h$ is a holomorphic section of a line bundle
over the complex structure moduli space. To preserve supersymmetry
the open membrane has to wrap the holomorphic 2-cycle $\Sigma_1$
of the CY. Though in the case of standard embedding without M5
branes the sum of (\ref{WOM}) over all curves $\Sigma_1$ in a
fixed homology class vanishes and likewise in the special cases of
non-standard embeddings which arise from weakly coupled heterotic
$(0,2)$ vacua related to linear sigma models, this is not the case
for a generic heterotic $(0,2)$ compactification with non-standard
embedding which we will assume in this paper.

Gaugino condensation occurs naturally on the hidden boundary
\cite{GC1} where the gauge theory becomes strongly coupled due to
the decrease of the deformed CY volume from visible to hidden
boundary (for $\beta_v>0$). It leads to a superpotential
\cite{GC2}
\beqa
W_{GC} = ge^{-\frac{1}{C_H}(S-\gamma(\cL)T )} \; ,
\label{WGC}
\eeqa
where
\beqa
\gamma(\cL) = \cG_v \cL \; ,
\qquad g = -C_H \mu^3 = -\frac{C_H}{32\pi^2}
\left(\frac{2 M_{GUT}}{M}\right)^3 = -5.0 \times 10^{-8}
C_H
\eeqa
and $C_H$ stands for the dual Coxeter number of the hidden gauge
group $H$. For instance for $H = E_8,\,E_6,\,SO(10),\,SU(5)$ one
has $C_H = 30,\,12,\,8,\,5$. The exponent receives an additive
contribution from the $T$ modulus as a result of the modified
gauge kinetic function for the hidden gauge group which is caused
by the non-trivial CV volume dependence on $\cL$. Note that the
fundamental 11d scale of heterotic M-theory is twice the grand
unified scale $M_{GUT}=3\times 10^{16}\,$GeV (`lowering of the
string scale') \cite{BD2} and therefore lower than the
conventional weakly coupled string scale. It is therefore
$2M_{GUT}$ which we have used above as an ultraviolet cut-off for
the gauge-theory. Note further that the appearance of small
numbers like $10^{-8}$ for $g$ is in part due to our conventions
whereby all scalar fields and superpotentials are dimensionless
and obtain their conventional mass dimensions through
multiplication by the appropriate power of $M=M_{Pl}/\sqrt{8\pi}$,
the reduced Planck mass. Likewise we expect similar small numbers
for $h$. As the canonical mass dimension of a superpotential is
$mass^3$ and we know that the highest available scale is
$2M_{GUT}$, the fundamental 11d scale, the absolute value for $h$
should have an upper bound of
\beqa
|h| \le (2M_{GUT}/M)^3 = 1.6\times 10^{-5}
\label{hBound}
\eeqa
in our conventions where $h$ becomes dimensionless by dividing
through $M^3$.

The `slope' $\gamma$ is controlled by the visible boundary
flux-parameter $\cG_v$ which is given through the integral
\beqa
\label{GFlux1}
\cG_v
= -\frac{l}{V_v}\left(\frac{\kappa}{4\pi}\right)^{2/3}
\int_{CY}\om\we\frac{(\tr F\we F -\frac{1}{2}\tr R\we R)_v}{8\pi}
\eeqa
over the visible boundary CY and $\om$ is the K\"ahler-form of the
undeformed CY. Using (\ref{Chern}), this parameter is related to
the visible instanton number $\beta_v$ in the following way
\beqa
\cG_v
= \frac{\pi l}{V_v}\left(\frac{\kappa}{4\pi}\right)^{2/3}
\beta_v\int_{\Sigma_1}\om \; .
\label{GFlux2}
\eeqa
For positive $\beta_v$ also $\cG_v$ will therefore be positive.
For $h^{(1,1)}=1$ the K\"ahler-form $\omega$ can be written in
terms of the single $H^{(1,1)}(X)$ basis-element $\omega_1$
\beqa
\omega = \left(\frac{6\cV_v}{d}\right)^{1/3} \omega_1
\eeqa
with the prefactor representing the single K\"ahler modulus. With
the estimate for the integral
\beqa
\int_{\Sigma_1}\om_1 \simeq V_v^{1/3}
\eeqa
which should be quite accurate for our case of $h^{(1,1)}=1$, and
the values (\ref{Units}) plus (\ref{Planck}) we obtain
\beqa
\cG_v
= \beta_v\Big(\frac{6}{d\cV_v}\Big)^{1/3}
\; .
\label{GFlux3}
\eeqa

\subsection{The Effective Potential}

The validity of the effective potential requires both $\cV$ and
$\cV_{OM}$ to be sufficiently larger than one such that multiply
wrapped instantons can be neglected. Furthermore, the derivation
of the K\"ahler-potential $K_{(C)}$ requires the expansion
parameter
\beqa
\label{Del}
\delta = \frac{H_{I\Jbar}C^I\overline{C}^{\Jbar}}{\cV_{OM}}
\eeqa
to be small. Notice that due to the employment of the full
background of \cite{CK1,CK3} we are not restricted to work in the
regime of $\epsilon\ll 1$ which via (\ref{Eps}) would generically
imply a hierarchy $\cV\gg\cV_{OM}$. Note, that this would mean
that gaugino condensation which is of $\cO(e^{-\cV/C_H})$ would be
drastically exponentially suppressed against the open membrane
instantons which go like $\cO(e^{-\cV_{OM}})$. However, in the
regime where $\epsilon=\cO(1)$ the contributions of gaugino
condensation and open membrane instantons are generically of same
size and therefore the possibility arises to balance both effects
against each other as we will see later.

As the 4d effective heterotic M-theory is a supergravity theory
with $N=1$ supersymmetry its potential for the moduli is given by
the standard formula
\beqa
U &= M^4 e^{K}\left(K^{\ibar j}D_{\ibar}\Wbar D_j W - 3 |W|^2
\right) + U_D
\label{Pot}
\eeqa
where $D_i W = \partial_i W + K_i W$ are the K\"ahler covariant
derivatives, $i,j$ run over all chiral fields and $M$ denotes the
reduced Planck mass. Moreover
\beqa
\label{Dterm}
U_D  = M^4 \frac{18\pi^2}{\cV\cV_{OM}^2}\sum_a \left( \Cbar T^a C
\right)^2
= M^4 \frac{18\pi^2}{\cV\cV_{OM}^2}\sum_a \left( H_{I\Jbar}
\Cbar^{\Jbar} {(T^a)^I}_K C^K \right)^2 \; .
\eeqa
is the D-term for the charged scalars. The $T^a,\; a
=1,\hdots,\text{dim} H$ are the generators of the unbroken visible
gauge group $H$.

Since $\delta,1/\cV,1/\cV_{OM}$ have to be considered as small, a
hierarchy is introduced among the terms appearing in the
potential. We will therefore find dominant contributions and terms
which are suppressed by at least one of these small entities. Let
us first concentrate on the open membranes alone and omit gaugino
condensation in (\ref{WNP}). Using the expressions collected in
the appendix one sees that the dominant terms come from the
$K^{\ibar j} \partial_{\ibar}\Wbar \partial_j W$ piece while the
$K^{\ibar j} \partial_{\ibar} \Wbar K_j W$ piece contributes only
via $K^{\Tbar T} \partial_{\Tbar} \Wbar K_T W$ to the dominant
term cubic in $C$. Moreover, the D-term contributes at this order.
All other terms which we will suppress in the following are
smaller by at least one additional factor of
$\delta,1/\cV,1/\cV_{OM}$. The dominant terms of the effective
potential are
\begin{alignat}{3}
U =\; &M^4e^K\left(K^{\ibar j}\partial_{\ibar}\Wbar \partial_j W
+2\Re(K^{\Tbar T}\partial_{\Tbar}\Wbar K_T W)\right)
+U_D  \\
=\; &\frac{M^4e^{K_{(A)}+K_{(Z)}}}{d\cV\cV^2_{OM}}
\bigg(\frac{|h|^2}{2}\cV_{OM}e^{-2\cV_{OM}}
+ \frac{3}{2}(1-\cN)
\Re\big(\hbarr e^{-\Tbar}\Lambda C^3\big)
+ \Big(\frac{3}{2}\Big)^2\cN\,|\Lambda C^2|^2 \bigg) \notag \\
&+ \frac{18\pi^2M^4}{\cV\cV_{OM}^2}\sum_a \big( \Cbar T^a C
\big)^2 \; .
\label{OMPot1}
\end{alignat}
where
\beqa
\label{Inequ}
\cN(\cL) = \frac{1}{\left(1+\beta_v\frac{2\cV_{OM}}{3\cV}\right)}
< 1
\eeqa
and we have used the compact notation
\begin{alignat}{3}
\Lambda C^3 &= \Lambda_{IJK} C^I C^J C^K  \\
|\Lambda C^2|^2 &= H^{\Ibar L} \Lambdabar_{\Ibar\Jbar\Kbar}
\Lambda_{LMN}\Cbar^{\Jbar}\Cbar^{\Kbar} C^M C^N \; .
\end{alignat}
The inequality (\ref{Inequ}) stems from the fact that we consider
$\beta_v>0$ as explained before. The single negative contribution
to the potential, $-3|W|^2$, is among the neglected suppressed
terms which means that the potential has to be {\em positive}.
That this is indeed true irrespective of the values of
$\delta,1/\cV,1/\cV_{OM}$ can be easily shown. Namely, by
investigating the suppressed terms, one finds that a $+|W|^2$ term
resp.~a $+3|W|^2$ term come from
\begin{alignat}{3}
\label{KSKS}
K^{\Sbar S} K_{\Sbar}\Wbar K_S W &=
\left(1+\beta_v\frac{H_{I\Jbar}C^I\Cbar^{\Jbar}}{\cV}\right)^2
|W|^2 \; , \\
\label{KTKT}
K^{\Tbar T} K_{\Tbar}\Wbar K_T W &=
3\left(1+\frac{H_{I\Jbar}C^I\Cbar^{\Jbar}}{2\cV_{OM}}\right)^2
|W|^2   \; .
\end{alignat}
which together overcompensate the negative $-3|W|^2$ contribution
and prove the positivity of the potential on the whole moduli
space. Actually this was to be expected in view of the `no-scale'
structure, $K_{(T)} = - 3\ln(T+\Tbar)+\hdots$, of the
K\"ahler-potential.

Let us next consider the full non-perturbative contribution to
(\ref{WNP}) comprising both open membrane instantons and gaugino
condensation. Again we will keep only the dominant terms in view
of the smallness of $\delta,1/\cV,1/\cV_{OM}$. With help of the
technical expressions of the appendix one finds that
\begin{alignat}{3}
K^{\ibar j}\partial_{\ibar}\Wbar \partial_j W =
&\;\frac{4\cV^2}{C_H^2}|W_{GC}|^2
+ \frac{4}{3}\cV_{OM}^2|W_{OM}-\frac{\gamma}{C_H}W_{GC}|^2 \\
&+ 4\cN\cV_{OM}\Re\Big(\big[-\Wbar_{OM}+\frac{1}{C_H}
(\gamma-2\beta_v)\Wbar_{GC}\big]\Lambda C^3\Big)
+ 6\cN\cV_{OM}|\Lambda C^2|^2  + \hdots        \notag
\end{alignat}
and
\begin{alignat}{3}
&\Re(K^{\Sbar S}\partial_{\Sbar}\Wbar K_S W
+ K^{\Tbar T}\partial_{\Tbar}\Wbar K_T W) \notag \\
=\,\;&\Re\Big(\big[ 2\cV_{OM}\Wbar_{OM}
+ \frac{4}{C_H}(\cV-\gamma\cV_{OM})\Wbar_{GC}\big]\Lambda C^3\Big)
+ \hdots
\end{alignat}
are the dominant terms while all other terms and those indicated
by dots are suppressed by at least a further factor
$\delta,1/\cV,1/\cV_{OM}$. Once more, one can verify that among
the suppressed terms, $K^{\Tbar T} K_{\Tbar} K_T |W|^2$,
contributes a $+3|W|^2$ which cancels the negative $-3|W|^2$ piece
in (\ref{Pot}). Since all other terms are manifestly positive, it
is clear that the effective potential has to be {\em positive}.
Again, positivity was to be expected from the fact that $K_{(T)}$
is of the form $-3\ln (T+\Tbar)+\hdots$ which implies a
cancelation of the negative $-3|W|^2$ contribution. If furthermore
$W$ would be independent of $T$ then the resulting tree level
cosmological constant coming from the $T$ sector would be zero.
Here, however, $W$ depends on $T$ which therefore generates a
further positive contribution to the potential.

At leading order in $\delta,1/\cV,1/\cV_{OM}$ the potential
becomes
\begin{alignat}{3}
U =\; &M^4e^K\left(K^{\ibar j}\partial_{\ibar}\Wbar \partial_j W
+2\Re(K^{\Sbar S}\partial_{\Sbar}\Wbar K_S W
+ K^{\Tbar T}\partial_{\Tbar}\Wbar K_T W)\right)
+U_D  \\
= \; &\frac{3M^4e^{K_{(A)}+K_{(Z)}}}{8d\cV\cV^2_{OM}}
\Bigg( \frac{4\cV^2}{C_H^2\cV_{OM}}|W_{GC}|^2
+\frac{4}{3}\cV_{OM}|W_{OM}-\frac{\gamma}{C_H}W_{GC}|^2 \notag
\\
+ \; &\Re\Big(4(1-\cN)\Wbar_{OM}\Lambda C^3
+ \frac{4}{C_H}
\big(\frac{\cV}{\cV_{OM}}-\gamma+\cN(\gamma-2\beta_v)\big)
\Wbar_{GC}\Lambda C^3
\Big)                              \label{OMGCPot1} \\
+ \; &6\cN|\Lambda C^2|^2\Bigg)
+ \frac{18\pi^2M^4}{\cV\cV_{OM}^2}\sum_a
\left( \Cbar T^a C \right)^2 + \hdots \; ,  \notag
\end{alignat}
where the dots indicate the omitted suppressed terms. Notice the
squares in the second line. Obviously, a trivial minimization of
the potential, in view of its positivity, is given by
$W_{GC}=W_{OM}=C=0$ which corresponds to the decompactification
vacuum characterized by $\cV_{OM},\cV\rightarrow\infty$. Our main
goal in this paper will be to show that there are further
non-trivial local minima of the potential with positive energy
corresponding to metastable de Sitter vacua.

\subsection{Volume Modulus and Parameter Range}

We also need to specify the dependence of $\cV(\cL)$ on $\cL$. In
the full non-linear background of \cite{CK1},\cite{CK3}, which we
use here, the deformed CY volume depends quadratically on the
orbifold coordinate $x^{11}$
\beqa
\label{Vol}
V(x^{11}) = V_v \left(1-\cG_v \frac{x^{11}}{l}\right)^2 \; .
\eeqa
One then infers, using (\ref{VLMod}), the following average CY
volume
\beqa
\label{AveVol}
\cV(\cL) = \cV_v\left(1 - \cG_v \cL
+\frac{1}{3}(\cG_v \cL)^2\right) \; .
\eeqa
$\cV(\cL)$ is monotonously decreasing throughout the interval
$\cL\in [0,\frac{3}{2\cG_v}]$. Since the moduli space for $\cL$ is
restricted to the smaller interval $[0,\frac{1}{\cG_v}]$, as we
will see shortly, $\cV(\cL)$ is monotonously decreasing with
$\cL$. Via (\ref{VOM}) also the complete $\cL$ dependence of
$\cV_{OM}(\cL)$ is now determined. In contrast to $\cV(\cL)$ one
finds that $\cV_{OM}(\cL)$ is monotonously increasing for all
values of $\cL$.

An interesting constraint which is imposed by the 11-dimensional
theory, results from the fact that at $x^{11}_0 = l/\cG_v$ a naked
singularity appears in the geometric background \cite{CK1},
\cite{CK3}. This necessitates the following upper bound on the
orbifold length modulus $\cL$
\beqa
\cL \le \cL_{max} = \frac{1}{\cG_v} \; .
\label{Lbound}
\eeqa
A stabilization of $\cL$ should therefore occur below or at
$\cL_{max}$. Indeed a stabilization close to $\cL_{max}$ would
result in a very precise prediction of the 4d Newton's Constant
\cite{WWarp},\cite{CK3} and is therefore clearly desirable. Notice
that the upper bound on $\cL$ leads directly to an upper bound on
$\gamma(\cL)$
\beqa
\gamma(\cL) \le 1 \; .
\label{gammabound}
\eeqa

Let us finally give for orientation the phenomenologically favored
visible boundary CY volume. The issue of how one might stabilize
this modulus willf be addressed in the final section. In the
phenomenological regime the CY radius on the visible boundary is
given by (\cite{BD1},\cite{CK3})
\beqa
\label{CYRadius}
V_v^{1/6} \simeq 2\kappa^{2/9} \; .
\eeqa
Together with (\ref{Units}) and (\ref{Planck}) this implies a
value $\cV_v \simeq 300$ for the dimensionless normalized volume.
Note however that this is merely a first order estimate and we
should therefore just take the order of magnitude for granted.

\section{Analysis of the Potential Including Open Membranes}

The prime idea which we will pursue here and in the following
sections is whether non-perturbative effects in conjunction with
$G_{lmnp}$ fluxes might generate a boundary-boundary
potential\footnote{For earlier investigations of boundary-boundary
potentials see \cite{KPot}.} which will cause a stabilization in
particular of the orbifold length (`dilaton'). The influence of
the fluxes at this stage comes from a deformation of the geometric
background which in turn enters the 4d effective potential through
the K\"ahler-potential and the geometric exponents of open
membrane instantons and gaugino condensation. Let us first
investigate the simplest case where gaugino condensation is absent
but open membrane instantons stretching from boundary to boundary
are included. The potential is given by (\ref{OMPot1}) and we will
now look for its extrema to see whether further de Sitter vacua
are present beyond the global minimum with vanishing energy
describing the decompactified vacuum.

\subsection{The Axion Sector}

To start the analysis of the potential (\ref{OMPot1}) for local
minima, it is most convenient to consider the $T$-modulus axion
$\sigma_T$ first. To this aim let us define
\beqa
V_1 = \Re\left(\hbarr\Lambda C^3\right) \; , \qquad V_2
= \Im\left(\hbarr\Lambda C^3\right)
\eeqa
such that we can write
\beqa
\Re\left(\hbarr e^{-\Tbar}\Lambda C^3\right)
= e^{-\cV_{OM}}\left(V_1\cos\sigma_T-V_2\sin\sigma_T\right) \; .
\eeqa
Extremization of the potential (\ref{OMPot1}) w.r.t.~the axion
$\sigma_T$, i.e.~setting $\partial U/\partial\sigma_T = 0$,
implies its fixation at
\beqa
\label{AxTMinOM}
\sigma_T = -\arctan\left(\frac{V_2}{V_1}\right) + n\pi \; ,
\eeqa
where $n\in\mathbf{Z}$. From this we get
\beqa
e^{i\sigma_T} = (-1)^n\frac{V_1-iV_2}{|V_1-iV_2|}
\eeqa
and finally obtain that at the extremal point
\beqa
\Re\left(\hbarr e^{-\Tbar}\Lambda C^3\right)
= (-1)^n e^{-\cV_{OM}}|V_1+iV_2| = (-1)^n |h| e^{-\cV_{OM}} |\Lambda C^3|
\; .
\eeqa
Having fixed $\sigma_T$ the potential at the extremum becomes
\begin{alignat}{3}
U =\; &\frac{M^4 e^{K_{(A)}+K_{(Z)}}}{d\cV\cV^2_{OM}}
\Bigg(\frac{|h|^2}{2}\cV_{OM}e^{-2\cV_{OM}}
+ \frac{3}{2}(-1)^n(1-\cN)
|h|e^{-\cV_{OM}}|\Lambda C^3|
+\Big(\frac{3}{2}\Big)^2\cN\,|\Lambda C^2|^2
\Bigg) \notag \\
&+ M^4 \frac{18\pi^2}{\cV\cV_{OM}^2}\sum_a
\big( \Cbar T^a C \big)^2 \; .
\label{OMPot2}
\end{alignat}

Notice that open membrane instantons alone cannot fix the axion
$\sigma_S$. This will become possible once we include gaugino
condensation in the next section. Note further, the general
feature that fixing an axion still allows for an arbitrary
discrete parameter choice, that of $n$. It is however only the
class of even or odd values for $n$ which lead to distinct
physics. Obviously, in view of the inequality $1\ge\cN$
(\ref{Inequ}), it is for non-vanishing $C$ the odd sector
\beqa
n \in 2\mathbf{Z}+1
\eeqa
which leads to lower potential energy.

\subsection{The Charged Matter Sector}

The general extremization condition which follows in the charged
matter sector from $\partial U/\partial C^N = 0$ is
\begin{alignat}{3}
&\frac{e^{K_{(A)}+K_{(Z)}}}{8d}
\bigg(
2\cN H^{\Ibar L}\overline{\Lambda_{IJK} C^J C^K} \Lambda_{LMN}C^M
+ (1-\cN)\Wbar_{OM}\Lambda_{NIJ} C^I C^J
\bigg)   \notag \\
+ \;&2\pi^2\sum_a \Big(H_{I\Jbar}\Cbar^{\Jbar}{(T^a)^I}_K C^K \Big)
H_{L\Mbar}\Cbar^{\Mbar} {(T^a)^L}_N
= 0  \; .
\label{CMinOM1}
\end{alignat}
These are as many complex equations as there are unknown complex
components of $C$. Therefore the solution to these equations will
fix $C$ completely. Obviously
\beqa
\label{CMinOM}
C_0 = 0
\eeqa
is one solution. We will however see soon that any non-trivial
solution for $C$ gives a lower potential energy than $C_0=0$.
Consequently the extremal value $C_0=0$ must correspond to a
maximum of the potential while one of the {\em non-trivial
solutions $C_0\ne 0$ gives the minimum of lowest energy}. This is
good news as it was apparently one of the unsolved open problems
of the weakly-coupled heterotic string to stabilize the $C$'s at
some non-vanishing values once supersymmetry was broken e.g.~by
gaugino condensation (see e.g.~\cite{RM}). Therefore the weakly
coupled heterotic string led to charged massless scalars which are
experimentally ruled out.

Let us now concentrate on the $C_0\ne 0$ solution. Since a
non-trivial vev for $C$ will be of direct phenomenological
relevance, we are now aiming at determining its size. For this
purpose, let us contract equation (\ref{CMinOM1}) with $C^N$ and
analyze the resulting single equation $C^N\partial U/\partial C^N
= 0$
\beqa
\frac{e^{K_{(A)}+K_{(Z)}}}{8d}
\bigg( 2\cN |\Lambda C^2|^2
+ (1-\cN) \Wbar_{OM} \Lambda C^3
\bigg)
+ 2\pi^2\sum_a(\Cbar T^a C)^2 = 0 \; .
\label{CMinOM2}
\eeqa
Its imaginary part
\beqa
\Im (\Wbar_{OM}\Lambda C^3) = 0
\eeqa
turns out to be equivalent to the condition (\ref{AxTMinOM})
resulting from the axion sector thus showing that the conditions
coming from the axion and the charged matter sectors are
compatible. By noticing that due to the fixing of the $\sigma_T$
axion in (\ref{AxTMinOM}) one has
\beqa
\Re(\Wbar_{OM}\Lambda C^3) = (-1)^n e^{-\cV_{OM}}|h\Lambda C^3|\;,
\eeqa
we obtain for its real part
\beqa
\frac{e^{K_{(A)}+K_{(Z)}}}{8d}
\bigg( 2\cN |\Lambda C^2|^2
+ (-1)^n (1-\cN) e^{-\cV_{OM}} |h\Lambda C^3|
\bigg)
+ 2\pi^2\sum_a(\Cbar T^a C)^2 = 0 \; .
\label{CMinOMReal}
\eeqa
In conformity with what we said before about the preference of the
odd $n$ axion sector, we see here that because $1\ge\cN$, we won't
get a non-trivial solution for $C$ if $n$ is even since then all
terms are individually positive. A non-trivial solution $C_0$
arises only in the $n$ {\em odd} sector when the cubic term
becomes negative. Henceforth we will concentrate on this sector.

\begin{figure}[t]
  \begin{center}
  \epsfig{file=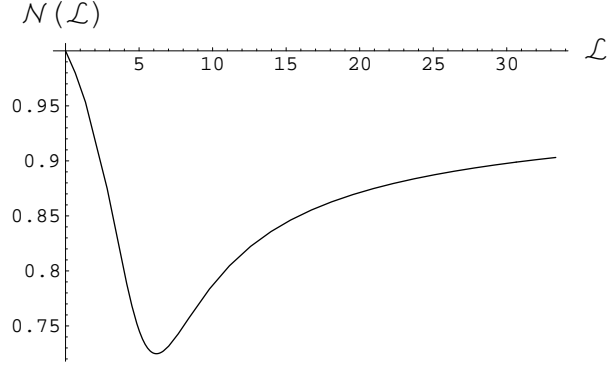,width=8cm,angle=0}
  \caption{\it The dependence of $\cN(\cL)$ on $\cL$ is plotted for
               the values $\beta_v=1,d=1,\cV_v=300$ which imply
               $\cG_v=0.3$. Notice that the physical range of $\cL$
               is limited and reaches from zero to
               $\cL_{max}=1/\cG_v=3.3$.}
  \label{calN}
  \end{center}
\end{figure}

To determine the size of the non-trivial $C_0$, we now focus on
the most influential factor, the exponential $e^{-\cV_{OM}(\cL)}$,
which amounts to study the $\cL$ dependence of $C_0$. Note that
the only $\cL$ dependence enters through $\cN(\cL)$ and
$\cV_{OM}(\cL)$. It is however clear that the polynomial $\cL$
dependence of $\cN(\cL)$ is much milder than the exponential $\cL$
dependence of $e^{-\cV_{OM}(\cL)}$. Indeed as illustrated in
figure \ref{calN} $\cN(\cL)$ differs only little from 1 so that
subsequently we will regard it as being constant $\cN=\cO(1)$.
Then only the cubic $C$ term exhibits a dependence on $\cL$
through the exponential and it is thus clear that satisfying
(\ref{CMinOMReal}) requires
\beqa
|C_0|\sim f e^{-\cV_{OM}(\cL)}
\eeqa
where $f$ does not depend on $\cL$. The size of the extremal $C_0$
will therefore be exponentially sensitive to $\cV_{OM}$ which is a
satisfactory result as it allows to bring $|C_0|$ easily down to
phenomenologically relevant values given that $f$ takes its
natural values at the reduced Planck scale (after reinstating
dimensions).

Let us now see how the potential looks like at the extremal point
when the extremization condition (\ref{CMinOMReal}) is applied.
With (\ref{CMinOMReal}) we can either eliminate all quartic $C$
terms in (\ref{OMPot2}) and replace them by a cubic term or vice
versa. The first choice leads to
\beqa
U = \frac{M^4 e^{K_{(A)}+K_{(Z)}}}{d\cV\cV^2_{OM}}
\bigg(\frac{|h|^2}{2}\cV_{OM}e^{-2\cV_{OM}} -
\frac{3}{8}(1-\cN)e^{-\cV_{OM}}|h||\Lambda C_0^3|
\bigg)  \; ,
\label{OMPot3}
\eeqa
which clearly shows that the non-trivial solutions $C_0\ne 0$ lead
to a lower potential energy than the trivial solution $C_0=0$
which gives maximal potential energy. It furthermore shows that
because the $C_0$ is proportional to $e^{-\cV_{OM}}$, that the
charged matter dependent part of the potential is suppressed at
the critical point by four times this exponential factor instead
of just two times like the $C_0$ independent open membrane part.
Furthermore the charged matter part is of lower order in
$1/\cV_{OM}$. We can therefore drop the $C_0$ dependent part for
the subsequent investigation of the critical point. The potential
at the critical point then becomes
\beqa
\label{OMPot3}
U =\; M^4 \frac{e^{K_{(A)}+K_{(Z)}}|h|^2}{2d}
\left(\frac{e^{-2\cV_{OM}(\cL)}}{\cV(\cL)\cV_{OM}(\cL)}\right)
\; .
\eeqa

\subsection{The Orbifold Length}

Let us finally see whether the potential exhibits a minimum along
the $\cL$ direction in moduli space and therefore allows for a
stabilization of the orbifold length. At the locus determined by
(\ref{AxTMinOM}) and (\ref{CMinOM}) where the potential becomes
extremal, the value of the potential is given by (\ref{OMPot3}).
Here, only the bracket in (\ref{OMPot3}) depends on $\cL$ such
that the extremization condition $U'=0$ (a prime denotes the
derivative w.r.t.~$\cL$) amounts to solving
\beqa
\label{LMin1}
\big(2\cV_{OM}+\ln(\cV_{OM}\cV)\big)'
= \frac{1}{\cL}(2\cV_{OM}+1)+\frac{2\cV'}{3\cV}(\cV_{OM}+2) = 0
\; ,
\eeqa
where to arrive at the second equation we have employed
(\ref{VOM}). Since we are working in the $\cV_{OM}\gg 1$ regime,
we can neglect both constants in brackets and obtain
\beqa
\label{LMin2}
3\cV + \cV' \cL = 0 \; .
\eeqa
This simplified equation is equivalent to setting ${\cV'}_{OM}=0$.
Substituting (\ref{AveVol}) for $\cV$, this equation becomes a
simple quadratic equation in $\cL$
\beqa
\cL^2 - \frac{12}{5\cG_v}\cL + \frac{9}{5\cG_v^2} = 0 \; .
\eeqa
This equation, however, has no real solution for $\cL$. Therefore,
for $\cV_{OM}\gg 1$ the minimization problem does not have a
solution which means that there is no further local vacuum in
addition to the global decompactification vacuum.

Let us note that for $\cV_{OM}\ll 1$ one can find non-trivial
solutions $\cL_0$ to the full equation (\ref{LMin1})
\cite{KSUSY03}. For example for an intersection number $d=1$ and
value $\cV_v=300$ one obtains solutions $\cL_0\ll 1$ provided that
$\beta_v\ge\cO(10^3)$. However, in this regime the hitherto
neglected terms with higher powers of $1/\cV_{OM}$ would have to
be added to the potential besides contributions of multiple
instanton wrappings which would then no longer be suppressed. To
avoid such complications and because length values $\cL\ll 1$ seem
to be outside the regime of validity of a local field theory
description, we will not consider this case further but instead
combine the open membrane instantons with the second naturally
appearing non-perturbative effect, gaugino condensation.

But before proceeding, a final comment on the background
dependence. It is evident from (\ref{LMin2}) that a background
which leaves (\ref{LMin2}) intact but provides a sufficiently more
negative derivative $\cV'$ than the one arising from
(\ref{AveVol}) would lead to a solution of (\ref{LMin2}). This is
actually the case for the linearized background of \cite{WWarp}
whose linear CY volume drops faster with $x^{11}$ than the full
quadratic volume (the former represents the tangent to the latter
at the location of the visible boundary) \cite{CK1},\cite{CK3}.
With the linearized background one finds indeed a minimum of the
potential at $\cL=\cL_{max}=1/\cG_v$ and therefore a stabilization
at maximal orbifold length. This analysis was carried out in
\cite{CK2} for the case with an additional 4d spacetime-filling M5
brane (the reason to include an M5 brane is to avoid the negative
CY volume problem arising within the linearized background, see
\cite{KSUSY03}). One then obtains from open membrane instantons an
$\cL$ dependence $U\sim e^{-\cV_{OM}}/(\cV\cV_{OM})$ instead of
$U\sim e^{-2\cV_{OM}}/(\cV\cV_{OM})$ for the case without M5 brane
which we face in this paper. In the large $\cV_{OM}$ limit,
however, any $U\sim e^{-a\cV_{OM}}/(\cV\cV_{OM})$, $a=const$,
leads to the same constraint (\ref{LMin2}). Therefore with or
without additional M5 brane, one obtains for the linearized
background a stabilization of $\cL$ at the maximally allowed
length because $\cV'$ is more negative in this background.
However, by going from the approximative (the approximation works
best close to the visible boundary and becomes worse the farther
one moves into the bulk) linearized background to the exact full
non-linear background, $\cV'$ becomes less negative and we cannot
find a solution to (\ref{LMin2}) any longer. Hence, open membrane
instantons aren't enough to stabilize $\cL$ in the regime of large
$\cV_{OM}$.

\section{Analysis of the Potential Including Open Membranes And
Gaugino Condensation}

\begin{figure}[bh]
  \begin{center}
  \epsfig{file=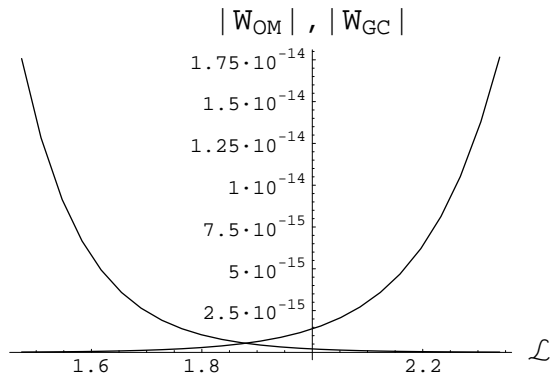,width=8cm,angle=0}
  \caption{\it The dependence of the absolute values of the open
               membrane and gaugino condensation superpotentials,
               $|W_{OM}|$ (left curve) and $|W_{GC}|$ (right
               curve) on $\cL$. $|W_{OM}|$
               decreases with $\cL$ while $|W_{GC}|$ increases steeply.
               For the plot we took the values $\beta_v=d
               =1, |h|=10^{-7}, \cV_v=300$ and hidden gauge
               group $SO(10)$, i.e.~$C_H=8$. The upper bound
               $\cL_{max}=1/\cG_v$ on $\cL$ lies at 3.7.}
  \label{Supos}
  \end{center}
\end{figure}

We will now include gaugino condensation (see \cite{GCRev} for a
recent review) as the second non-perturbative effect which arises
naturally in the strongly coupled gauge theory on the hidden
boundary. The main idea for the stabilization of $\cL$ is now the
following. From the absolute values of the superpotentials
\beqa
|W_{OM}| \sim |h|e^{-\cV_{OM}} \; ,
\qquad |W_{GC}| \sim |g|e^{-\frac{1}{C_H}(\cV-\gamma \cV_{OM})}
\eeqa
and the monotony properties of $\cV$ and $\cV_{OM}$, as pointed
out after (\ref{AveVol}), it is clear that $W_{OM}$ decreases with
$\cL$ while $W_{GC}$ increases in the relevant regime $\cL\in
[0,\frac{1}{\cG_v}]$ (see figure \ref{Supos}). Therefore one might
expect that also at the level of the potential (\ref{OMGCPot1})
the combination of both contributions could lead to a non-trivial
minimum due to the opposite monotony properties of the two
non-perturbative effects. We will now show that this balancing of
open membrane instantons
against gaugino condensation indeed works.

\subsection{The Axion Sector}

Starting with the potential (\ref{OMGCPot1}) let us first derive
its minimization constraints w.r.t.~the axions
$\sigma_S,\sigma_T$. In order to focus on the $\sigma_S,\sigma_T$
dependence of the potential, it proves useful to define the
following complex valued quantities
\begin{alignat}{3}
X_1+iX_2 &= e^{-\frac{\cV}{C_H}+(\frac{\gamma}{C_H}-1)\cV_{OM}} g\hbarr \\
Y_1+iY_2 &= e^{-\cV_{OM}}\hbarr \Lambda C^3 \\
Z_1+iZ_2 &= e^{-\frac{\cV}{C_H}+\frac{\gamma}{C_H}\cV_{OM}} g
\Lambda C^3 \; .
\end{alignat}
The portion of the potential which depends on $\sigma_S,\sigma_T$
is captured by the three mixed terms
\begin{alignat}{3}
\Re(\Wbar_{OM} W_{GC}) &=
X_1\cos\Big(\frac{\sigma_S}{C_H}-(\frac{\gamma}{C_H}+1)\sigma_T\Big)
+ X_2\sin\Big(\frac{\sigma_S}{C_H}-(\frac{\gamma}{C_H}+1)\sigma_T\Big)
\label{Mix1} \\
\Re(\Wbar_{OM}\Lambda C^3) &= Y_1\cos\sigma_T - Y_2\sin\sigma_T \\
\Re(\Wbar_{GC}\Lambda C^3) &=
Z_1\cos\Big(\frac{\sigma_S}{C_H}-\frac{\gamma}{C_H}\sigma_T\Big) -
Z_2\sin\Big(\frac{\sigma_S}{C_H}-\frac{\gamma}{C_H}\sigma_T\Big)
\; ,
\end{alignat}
where the first term arises from expanding the complete square
$|W_{OM}-\frac{\gamma}{C_H} W_{GC}|^2$. The extremization
conditions of the potential w.r.t.~$\sigma_S$ and $\sigma_T$ are
$\partial U/\partial\sigma_S = \partial U/\partial\sigma_T = 0$.
However more succinct and easier to work with are the equivalent
conditions, $\partial U/\partial\sigma_S - \partial
U/\partial\sigma_T = 0$ which reads
\begin{alignat}{3}
&2\frac{\gamma}{C_H}\cV_{OM} \left(
X_1\sin\Big(\frac{\sigma_S}{C_H}-(\frac{\gamma}{C_H}+1)\sigma_T\Big)
- X_2 \cos
\Big(\frac{\sigma_S}{C_H}-(\frac{\gamma}{C_H}+1)\sigma_T\Big)
\right)
\notag \\
= \;&3(\cN-1)
\left(Y_1\sin\sigma_T+Y_2\cos\sigma_T\right) \; ,
\label{AxMinOMGC1}
\end{alignat}
and $(\gamma + C_H)\partial U/\partial\sigma_S - \gamma \partial
U/\partial\sigma_T$ which is
\begin{alignat}{3}
&\frac{1}{C_H}\Big(\frac{\cV}{\cV_{OM}}-\gamma+\cN(\gamma-2\beta_v)\Big)
\left(Z_1\sin\Big(\frac{\sigma_S}{C_H}-\frac{\gamma}{C_H}\sigma_T\Big)
+ Z_2\cos\Big(\frac{\sigma_S}{C_H}-\frac{\gamma}{C_H}\sigma_T\Big)
\right) \notag \\
= \;&(\cN-1)
\left(Y_1\sin\sigma_T+Y_2\cos\sigma_T\right) \; .
\label{AxMinOMGC2}
\end{alignat}
These two equations fix the axions $\sigma_S,\sigma_T$ in terms of
$\cL$ and the charged scalars. Alternatively they can also be
expressed in terms of the original variables as
\begin{alignat}{3}
-&\frac{2}{3} \frac{\gamma}{C_H} \cV_{OM} \Im\left(\Wbar_{OM}
W_{GC}\right)
= (\cN-1) \Im\left(\Wbar_{OM}\Lambda C^3\right) \notag \\
= \;&\frac{1}{C_H}\Big(\frac{\cV}{\cV_{OM}}-\gamma+\cN(\gamma-2\beta_v)
\Big) \Im\left(\Wbar_{GC}\Lambda C^3\right)
\label{AxMinOMGC3}
\end{alignat}

\subsection{The Charged Matter Sector}

In the charged scalar sector the extremization constraint,
$\partial U/\partial C^N = 0$, can straightforwardly be evaluated
to give
\begin{alignat}{3}
&\frac{e^{K_{(A)}+K_{(Z)}}}{8d}
\bigg(
2\cN H^{\Ibar L}\overline{\Lambda_{IJK}C^JC^K} \Lambda_{LMN}C^M
+ \frac{1}{C_H}\Big(\frac{\cV}{\cV_{OM}}-\gamma+\cN(\gamma-2\beta_v)
\Big)\Wbar_{GC}\Lambda_{NIJ}C^IC^J   \notag \\
+ \;&(1-\cN)\Wbar_{OM}\Lambda_{NIJ}C^IC^J
\bigg)
+ 2\pi^2\sum_a \Big(H_{I\Jbar}\Cbar^{\Jbar}{(T^a)^I}_K C^K\Big)
H_{L\Mbar}\Cbar^{\Mbar} {(T^a)^L}_N
= 0  \; .
\label{CMinOMGC1}
\end{alignat}
These are as many complex equations as there are unknown complex
components of $C$. The solution to this system of equations will
therefore fix all components of $C$.

A more handy, and for our purposes sufficient, implied constraint
consists of the index-free contracted equation $C^N\partial
U/\partial C^N = 0$ which reads
\begin{alignat}{3}
&\frac{e^{K_{(A)}+K_{(Z)}}}{8d}
\bigg( 2\cN |\Lambda C^2|^2
+ (1-\cN) \Wbar_{OM} \Lambda C^3
+ \frac{1}{C_H}\Big(\frac{\cV}{\cV_{OM}}-\gamma+\cN(\gamma-2\beta_v)
\Big) \Wbar_{GC}\Lambda C^3
\bigg) \notag \\
+ &2\pi^2\sum_a(\Cbar T^a C)^2 = 0 \; .
\label{CMinOMGC2}
\end{alignat}
It is obvious that
\beqa
C_0 = 0
\eeqa
constitutes a solution not only to the contracted equation but
also to the full set of constraints, $\partial U/\partial C^N
= 0$. There is however, as pointed out, also a {\em nontrivial
solution $C_0$}. As already indicated before, this is interesting
as it solves the notorious problem of vanishing $C$'s, and
therefore massless charged scalars, of the weakly coupled
heterotic string after supersymmetry breaking through gaugino
condensation \cite{RM}. Let us now investigate further the
constraint (\ref{CMinOMGC2}) for the non-trivial $C$.

Observe first that in the complex valued equation
(\ref{CMinOMGC2}) only the second and third term possess an
imaginary part, so that the real valued equation resulting from
just the imaginary part of the complex equation (\ref{CMinOMGC2})
simply becomes equal to the second equation in (\ref{AxMinOMGC3})
or equivalently equal to (\ref{AxMinOMGC2}). It therefore
represents no further constraint beyond those already obtained
from the axion sector but shows that the extremization conditions
coming from the axion sector and the charged matter sector are
compatible.

On the other hand the real part of the equation $C^N\partial
U/\partial C^N = 0$ gives us the independent constraint
\begin{alignat}{3}
&\frac{e^{K_{(A)}+K_{(Z)}}}{8d} \bigg( 2\cN |\Lambda C^2|^2
+ (1-\cN)\Re\left(\Wbar_{OM}\Lambda C^3\right)
+ \frac{1}{C_H}\Big(\frac{\cV}{\cV_{OM}}-\gamma+\cN(\gamma-2\beta_v)\Big)
\notag \\
&\times\Re\left(\Wbar_{GC}\Lambda C^3\right)
\bigg)
+ 2\pi^2\sum_a(\Cbar T^a C)^2
= 0  \; .
\label{CMinOMGCReal}
\end{alignat}
This is an important equation because it controls the size of
$C_0$ and therefore the size of the $C$ dependent terms in the
potential at the extremal point. This in turn will show whether
the $C$ dependent terms are of the same order (or bigger) than the
sofar included $C$ independent dominant terms and therefore have
to be kept in the analysis or whether the $C$ dependent terms are
of the same small size as the suppressed terms which would mean
that the $C$ terms likewise would have to be discarded.

Since we have used $1/\cV$ and $1/\cV_{OM}$ as expansion
parameters (only sufficiently large $\cV$ and $\cV_{OM}$ allow us
to neglect contributions from multiple instanton wrappings which
are currently not well understood) to determine the leading
dominant part of the potential, the question will therefore be how
does $C_0$ depend on these two parameters. Let us remark that
though $\cN(\cL)$ has a dependence on these two parameters it is
with sufficient accuracy constant, $\cN=\cO(1)$, (see figure
\ref{calN}) and is therefore of no concern in the sequel. If we
furthermore approximate the real parts in (\ref{CMinOMGCReal})
through absolute values then the magnitude of $C_0$ is determined
by
\begin{alignat}{3}
f_1 |C_0|^4 + f_2 |h| e^{-\cV_{OM}}|C_0|^3
+ f_3 \frac{|g|}{C_H}
\Big(\frac{\cV}{\cV_{OM}}-\gamma+\cN(\gamma-2\beta_v)\Big)
e^{-\frac{1}{C_H}(\cV-\gamma\cV_{OM})}|C_0|^3 \sim 0 \notag \\
\end{alignat}
where the coefficients $f_i$ do not depend on $\cV$ and
$\cV_{OM}$. The size of $C_0$ is therefore estimated as
\beqa
|C_0|\sim \frac{f_2}{f_1} |h| e^{-\cV_{OM}}
+ \frac{f_3|g|}{f_1 C_H}
\Big(\frac{\cV}{\cV_{OM}}-\gamma+\cN(\gamma-2\beta_v)\Big)
e^{-\frac{1}{C_H}(\cV-\gamma\cV_{OM})} \; .
\label{CApprox}
\eeqa
The non-perturbative exponential factors are phenomenologically
interesting as they allow to bring $C$ down to relevant values far
below the reduced Planck scale (after reinstating dimensions for
$C$ by multiplying with $M$).

After having estimated the $C$ vev, let us now look at the
resulting vacuum energy. The condition (\ref{CMinOMGCReal})
implies, by using it to eliminate the two cubic $C^3$ terms, that
the potential (\ref{OMGCPot1}) at the critical point becomes
\begin{alignat}{3}
U = \; &\frac{3M^4e^{K_{(A)}+K_{(Z)}}}{2d\cV\cV^2_{OM}}
\Bigg( \frac{\cV^2}{C_H^2\cV_{OM}}|W_{GC}|^2
+\frac{1}{3}\cV_{OM}|W_{OM}-\frac{\gamma}{C_H}W_{GC}|^2
- \frac{1}{2}\cN|\Lambda C^2|^2 + \hdots
\Bigg) \notag \\
- \; &\frac{6M^4\pi^2}{\cV\cV_{OM}^2}\sum_a(\Cbar T^a C)^2 \; ,
\label{OMGCPot2}
\end{alignat}
where the dots comprise the neglected suppressed terms. It is
therefore clear from the negative definiteness of the quartic
$C^4$ terms that the {\em solution with non-vanishing $C_0$} is
the one with the {\em lower energy density} and hence preferred
over the $C_0=0$ solution. Notice however that though the quartic
terms enter with a negative sign, the total potential energy
density must still be non-negative in view of the cancelation of
the $-3|W|^2$ term, as has been shown earlier.

It is easy to see that for the further analysis we can now drop
the $C$ contribution. The $C^4$ terms of the potential at the
extremum are of magnitude
\beqa
|C_0|^4 \sim
\Big( |h|e^{-\cV_{OM}}
+ \frac{|g|\cV}{C_H\cV_{OM}} e^{-\frac{1}{C_H}(\cV-\gamma\cV_{OM})}
\Big)^4 \; .
\eeqa
They are therefore strongly suppressed against the pure
non-perturbative contributions not only because they are of lower
order in $1/\cV,1/\cV_{OM}$ but also because they are suppressed
by the fourth power of the exponentials while the non-perturbative
contributions go like the second power of the exponentials. We can
therefore conclude that though a non-trivial vev for $C$ is a
general outcome of our analysis, the $C$ dependent terms can
safely be neglected at the critical point in comparison to the
non-perturbative contributions. They will consequently be omitted
henceforth.

Before doing so, however, let us briefly comment on the by now
ubiquitous exponential factors and their virtues. With the $C^4$
terms exponentially suppressed as well, the whole potential
exhibits an exponential suppression at the critical point.
Therefore, at least at tree level, the resulting positive vacuum
energy might be brought down to a phenomenologically acceptable
value in the range of $U\sim \text{meV}^4$ triggered by values for
$\cV,\cV_{OM}$ of $\cO(100)$. In this respect, a hidden gauge
group of small rank with smaller value for the dual Coxeter number
$C_H$ will be favored. This idea of obtaining a small cosmological
constant through an exponential suppression was introduced earlier
in the string-theory brane-world context \cite{K3}. However while
for those brane-worlds an exponential warp-factor was exploited it
is here the non-perturbative effects which give the exponential
suppression. It would of course be interesting to investigate
quantum corrections to our effective potential and to see whether
they likewise come out exponentially suppressed. On the other hand
with values of $\cV/C_H,\cV_{OM}\sim\cO(10)$ one would get $C$
vev's and therefore masses at the TeV scale (see
e.g.~\cite{PolBook}). We will see later when we are studying the
values of $\cV,\cV_{OM}$ for our vacua that it is actually the
latter possibility which can be realized here. An explanation of
the tiny value of the observed cosmological constant is however
out of reach. Presumably the solution to this most puzzling enigma
seems to require radical new ideas about the structure of
spacetime itself which might perhaps be based on theories with
just a finite number of degrees of freedom as suggested by
\cite{dS1},\cite{dS3},\cite{BHE}.

\subsection{The Orbifold Length}

As explained we will henceforth omit the $C$ terms. As far as the
derivation of the axion constraints is concerned omitting the $C$
terms amounts to setting $Y_1=Y_2=Z_1=Z_2=0$ which means that the
initially two constraints (\ref{AxMinOMGC1}), (\ref{AxMinOMGC2})
collapse to just one constraint which fixes the linear combination
$a\sigma_S-(a\gamma+1)\sigma_T$ to be
\beqa
\frac{\sigma_S}{C_H}-\big(\frac{\gamma}{C_H}+1\big)\sigma_T
= \arctan\bigg(\frac{\Im \hbarr}{\Re \hbarr}\bigg) + n\pi
\; , \;\; n\in\mathbf{Z} \; .
\label{AxFix}
\eeqa
We will see below that its real complement $\frac{1}{C_H}\Re
S-(\frac{\gamma}{C_H}+1)\Re T$ determines the extremal value of
$\cL$ by becoming approximately zero. It is therefore the
combination of chiral fields $\frac{1}{C_H}S - (\frac{\gamma}{C_H}
+ 1)T$ which becomes fixed at the extremum at leading order.

Let us now actually perform the extremization w.r.t.~$\cL$.
Without $C$'s the leading potential simplifies to
\beqa
U = \frac{M^4 e^{K_{(A)}+K_{(Z)}}}{2d\cV\cV_{OM}}
\bigg(3\Big(\frac{\cV}{C_H\cV_{OM}}|W_{GC}|\Big)^2
+\Big|W_{OM}-\frac{\gamma}{C_H} W_{GC}\Big|^2\bigg)
\label{OMGCPot3}
\eeqa
Expanding the second square gives a mixed term which, with help of
(\ref{Mix1}) and the axion fixation (\ref{AxFix}), simplifies to
\beqa
\Re(\Wbar_{OM}W_{GC}) = (-1)^n |W_{OM}| |W_{GC}| \; .
\label{MixedTerm}
\eeqa
Consequently the potential can be written as
\begin{alignat}{3}
U &= \frac{M^4 e^{K_{(A)}+K_{(Z)}}}{2d\cV\cV_{OM}}
\bigg(3\Big(\frac{\cV}{C_H\cV_{OM}}|W_{GC}|\Big)^2
+\Big(|W_{OM}|-(-1)^n\frac{\gamma}{C_H} |W_{GC}|\Big)^2\bigg) \; .
\label{OMGCPot4}
\end{alignat}
Obviously the axion sector with even $n$
\beqa
n \in 2\mathbf{Z}
\eeqa
results in the lower potential energy density. Hence we will
concentrate on this even sector hereafter. The extremization
condition $\partial U/\partial\cL=0$ leads to the transcendental
equation
\beqa
\frac{\partial}{\partial \cL}
\bigg[
\frac{1}{\cV\cV_{OM}}
\bigg(3\Big(\frac{|g|\cV}{C_H\cV_{OM}}
e^{-\frac{\cV}{C_H}+\frac{\gamma\cV_{OM}}{C_H}}\Big)^2
+ \Big(|h|e^{-\cV_{OM}}
- \frac{|g|\gamma}{C_H}
e^{-\frac{\cV}{C_H}+\frac{\gamma\cV_{OM}}{C_H}}
\Big)^2
\bigg)
\bigg] = 0
\label{LMinOMGC}
\eeqa
which we have to solve numerically. We will now discuss its
solutions.

\subsection{De Sitter Minima}

\begin{figure}[t]
  \begin{center}
  \epsfig{file=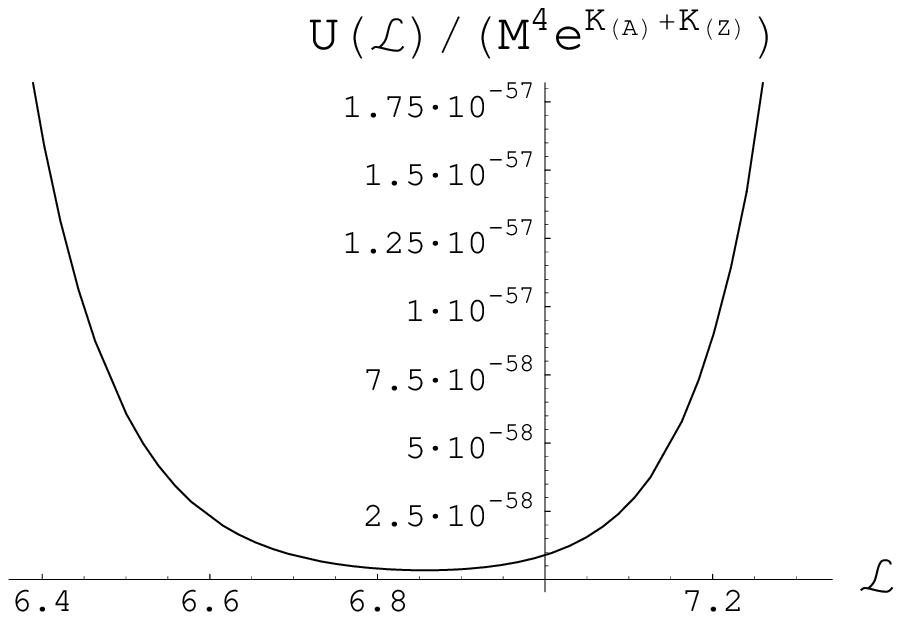,width=7cm,angle=0}
  \epsfig{file=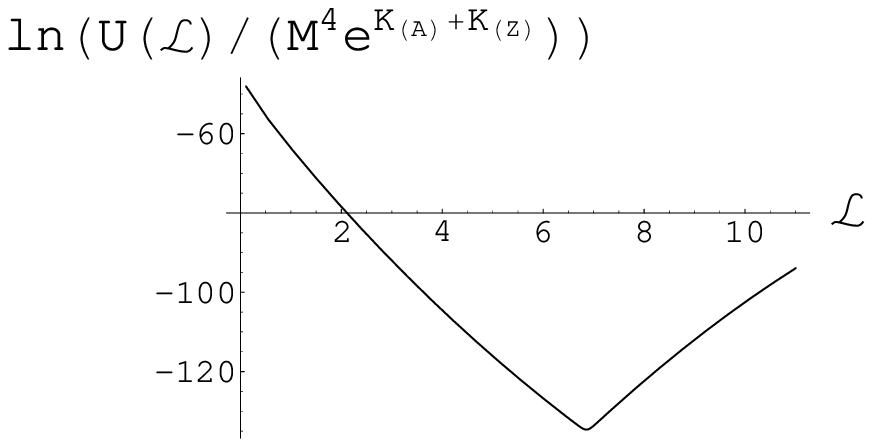,width=7cm,angle=0}
  \caption{\it The left picture shows the leading order effective
  4-dim.~potential in Planck units as a function
  of the orbifold length $\cL$. To give an impression of the global
  behavior of the potential over the complete interval
  $[0,\cL_{max}]$, we display in the right picture
  the logarithm of the potential. A de Sitter minimum appears at
  $\cL_0 = 6.9$. Towards its left open membrane instantons
  dominate the potential, towards its right it's the gaugino
  condensation. For the assumed values
  $\beta_v=1, d=10, |h|=10^{-8}, \cV_v=800$ and a hidden gauge group
  $SO(10)$ the upper bound on $\cL$ is $\cL_{max}=11.0$.}
  \label{OMGCPotential}
  \end{center}
\end{figure}

The first important result is that without the need for a
fine-tuning of the parameters generically an extremal value
$\cL_0$, i.e.~a real valued solution to this equation, is found
below the upper bound $\cL_{max}=1/\cG_v$ (see
tables~\ref{aTable}-\ref{VTable} in appendix~\ref{Tables}). That
this extremal value indeed corresponds to a {\em minimum} of the
potential in $\cL$ direction becomes apparent once the potential
(\ref{OMGCPot4}) for even $n$ is plotted. Figure
\ref{OMGCPotential} shows the potential for the case of a hidden
$SO(10)$.
Notice once more that the smallness of $|h|$ is natural in our
conventions, cf.~(\ref{hBound}). In order to capture the global
behavior of the potential in the whole admissible interval
$\cL\in[0,\cL_{max}]$ we have plotted in addition the logarithm of
the potential.
\begin{figure}[h]
  \begin{center}
  \epsfig{file=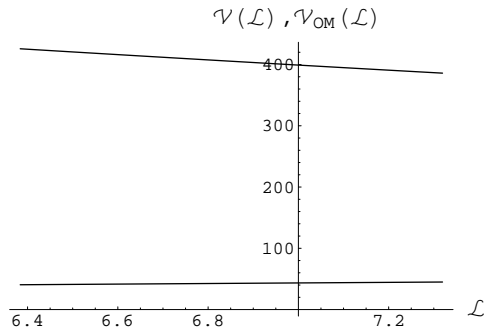,width=7cm,angle=0}
  \caption{\it To check that the minimum lies in a controllable
  regime we present here $\cV$ (upper curve) and $\cV_{OM}$ (lower
  curve) for the same parameter values as in the previous figure.
  Both are substantially larger than 1 at the minimum's position
  $\cL_0 = 6.9$ to guarantee that the neglected terms or
  multiply wrapped instantons are adequately suppressed.}
  \label{VVOM}
  \end{center}
\end{figure}
Characteristic in this second plot is the nearly linear dependence
both to the left and to the right of the minimum. It clearly
reveals the dominance of the open membrane instantons to the left
and that of gaugino condensation to the right with their different
monotonicity properties. To directly check that the minimum lies
in a regime where the theory is under control, we have plotted in
figure \ref{VVOM} the values for $\cV,\cV_{OM}$ around the
minimum. Indeed both volumes are sufficiently larger than 1 so
that one can trust the leading order terms included in the
potential and moreover can be sure that multiply wrapped instanton
contributions are considerably suppressed and need not be
considered.

As we see from figure \ref{OMGCPotential} and was also clear from
the fact that the $-3|W|^2$ term canceled out of the potential,
the minimum comes with a positive vacuum energy density and
represents hence a local {\em de Sitter vacuum}. It is easy to see
that at leading order in $1/\cV$ and $1/\cV_{OM}$
\beqa
D_S W = -\frac{W_{GC}}{C_H} \ne 0 \; ,
\eeqa
thus showing that at the location of the minimum in particular
$D_SW$ is non-vanishing. Consequently {\em supersymmetry is broken
spontaneously in these vacua} through F-term expectation values.

One might wonder whether our potential can be directly
extrapolated to the weakly coupled regime. This is however not
possible as it stands because we derived the potential in the
regime where both $\cV\gg 1,\cV_{OM}\gg 1$, in order to have the
supergravity theory under control. To connect to the weakly
coupled heterotic string one would have to go instead to the limit
where $\cV_v\sim \cV \gg 1$ but $\cV_{OM}\rightarrow 0$. In this
limit some terms which we have suppressed would become dominant
while some of those terms which we included would become
suppressed. Therefore the potential plotted in figure
\ref{OMGCPotential} should not be trusted far below the minimum
$\cL_0$ where one reaches the weakly coupled limit
$\cL_0\gg\cL\rightarrow 0$.

\subsection{How Robust is the Minimum?}

Let us now analyze how robust the mechanism to generate these de
Sitter vacua is. To this end we have to study the influence of the
choice of the hidden gauge group $H$, the choice of the
intersection number $d$ and therefore the choice of the CY.
Furthermore, since the numerical value of $|h|$ cannot be
calculated precisely with current technology, we will have to
investigate its influence as well. Finally, it will be important
to check that both volumes $\cV_0=\cV(\cL_0)$ and
$\cV_{OM,0}=\cV_{OM}(\cL_0)$ are substantially bigger than 1 at
the minimum's position as this secures that the minimum lies in a
region of moduli space which is under control in the sense that we
can trust the leading terms considered in the potential and
moreover do not have to worry that multiply wrapped instantons
would generate substantial corrections. Moreover from a
phenomenological perspective it will be interesting to see whether
we can achieve to bring the minimum's position $\cL_0$ close to
the maximum $\cL_{max}$ as this position is distinguished by
leading to the correct 4-dimensional Newton's Constant once the
GUT-scale and the GUT gauge coupling attain their traditional
values \cite{WWarp},\cite{CK3}.

Finally we aim to show that the minimum's position $\cL_0$ which
is obtained as a solution to the fairly complicated transcendental
equation (\ref{LMinOMGC}) can reasonably well be approximated by
$\cL_{prox}$ which is obtained from the much simpler equation
\beqa
\Big(\frac{1}{C_H}\Re S - \big(\frac{\gamma}{C_H}+1\big)\Re
T\Big) \Big|_{\cL=\cL_{prox}} = 0 \; .
\label{SimpleLMin}
\eeqa
This condition can be obtained from the full extremization
condition (\ref{LMinOMGC}) by neglect of the first term against
the second and by neglect of the derivative of a polynomial in
$\cV,\cV_{OM}$ against the derivative of an exponential in
$\cV,\cV_{OM}$. The former is justified as it is the second term
whose derivative changes sign near the minimum while the first
term's derivative doesn't change sign. The physics which is
captured by the simplified condition (\ref{SimpleLMin}) is the
{\em balance between open membrane instantons $(\sim e^{-\Re T})$
and gaugino condensation} $(\sim e^{-\frac{1}{C_H}\Re
S+\frac{\gamma}{C_H}\Re T})$ since it is equivalent to setting
\beqa
e^{-\Re T} = e^{-\frac{1}{C_H}\Re S+\frac{\gamma}{C_H}\Re T}
\eeqa
at $\cL=\cL_{prox}$. The so obtained approximate minimum's
position $\cL_{prox}$ does not depend on $|h|,|g|$ and represents
the true position most faithfully if $|h|\simeq |g|$.

Tables~\ref{aTable}-\ref{VTable} of appendix~\ref{Tables} display
the dependence of $\cL_0,\cL_{prox},\cL_{max},\cV_0,\cV_{OM,0}$ on
the hidden gauge group $H$ and $d,|h|,\cV_v$. Table~\ref{aTable}
which investigates the influence of the hidden gauge group $H$
clearly shows that in order to bring $\cL_0$ close to $\cL_{max}$
one would like to take a non-trivial hidden gauge bundle which
breaks the hidden $E_8$ gauge symmetry down to a group with
considerably smaller dual Coxeter number $C_H$. Furthermore, we
discern from table~\ref{aTable} that a hidden gauge group with low
$C_H$ will bring $\cV_{OM,0}$ up while maintaining rather large
values for $\cV_0$ thereby lowering both the vev for the charged
matter $C$ and the positive vacuum energy at the minimum.
Table~\ref{dTable} studies the influence of varying the CY
intersection number $d$. A larger $d$ drives $\cL_0$ to larger
values. This can be quantified by observing that a changing $d$
does not alter $\cV_0$ or $\cV_{OM,0}$. It is then clear from
(\ref{VOM}) that
\beqa
\cL_0 \propto d^{1/3} \; .
\eeqa
Finally table~\ref{hTable} and table~\ref{VTable} which study the
influence of $|h|$ and $\cV_v$ show that the influence of $|h|$ on
the position $\cL_0$ of the critical point is rather mild while
increasing values of $\cV_v$ let $\cL_0$ and $\cV_0,\cV_{OM,0}$
grow.

Moreover, we see from the tables of appendix \ref{Tables} that no
fine-tuning needs to be invoked to get both $\cV_0\gg 1$ and
$\cV_{OM,0}\gg 1$ in order to have the minimum located in a
controllable regime. Besides that we also see that $\cL_{prox}$
approximates $\cL_0$ rather well. In conjunction with the
stabilization of the axions (\ref{AxFix}) we hence see that at
leading order and under the approximation leading to
(\ref{SimpleLMin}) it is
\beqa
\frac{1}{C_H}S-\big(\frac{\gamma}{C_H}+1\big)T = i
\Big(\arctan\bigg(\frac{\Im \hbarr}{\Re \hbarr}\bigg) + 2\pi
n \Big)
\label{STFix}
\eeqa
which gets stabilized. Since $\partial U/\partial\cL=0$ is a
scalar equation it is of course clear that it cannot stabilize
both $\Re S$ and $\Re T$ individually. To stabilize both $\Re S$
and $\Re T$ individually requires a stabilization of $\cV_v$ as
well for which we will have to take into account also the $H$-flux
superpotential as will be outlined in the final section. Note,
however, that at subleading order we know that there are two
independent conditions (\ref{AxMinOMGC1}) and (\ref{AxMinOMGC2})
of the axion sector which stabilize both $\Im S$ and $\Im T$.

\subsection{Properties of the De Sitter Vacua}

Let us now look at some of the properties of our de Sitter vacua.
As mentioned before one can easily verify that the {\em $N=1,D=4$
supersymmetry is broken spontaneously} in them. At leading order
in $1/\cV$ and $1/\cV_{OM}$ one finds
\begin{alignat}{3}
D_S W &= -\frac{1}{C_H}W_{GC} \ne 0 \\
D_T W &= -W_{OM}+\frac{\gamma}{C_H}W_{GC} \\
D_I W &\equiv D_{C^I} W
= 3\Lambda_{IJK}C^JC^K
+ \frac{3}{2\cV_{OM}\cN}H_{I\Jbar}\Cbar^{\Jbar} \; .
\end{alignat}
Because of (\ref{STFix}) we can write for these vacua
\beqa
D_T W
= \Big(\frac{\gamma}{C_H}\left|\frac{g}{h}\right|-1\Big) W_{OM} \; .
\label{DTW}
\eeqa
Due to the balancing of open membrane instantons and gaugino
condensation at the potential's minimum, $D_S W$ and $D_T W$ will
be roughly of the same order. As to $D_I W$, its first term will
in view of (\ref{CApprox}) be exponentially smaller than the
second term and is therefore negligible while the second term is
suppressed by an additional $1/\cV_{OM}$ as compared to $D_S W$
and $D_T W$. One therefore finds in these vacua (except for the
fine-tuned case where the bracket in (\ref{DTW}) vanishes and one
has $D_T W=0$)
\beqa
|D_S W|\sim |D_T W| \gg |D_I W| > 0 \; ,
\eeqa
and therefore an {\em F-term supersymmetry breaking}. The last
inequality holds because of the non-vanishing $C$'s in these
vacua. Note that for $C\ne 0$ also the D-term might be
non-vanishing. However because it contains four $C$'s it would be
strongly exponentially suppressed due to (\ref{CApprox}) against
the F-terms.

The next important characterization of our vacua will then be the
supersymmetry breaking scale $M_{SUSY}$. It is given by the
largest vev of the F-terms
\beqa
M_{SUSY} = M \, \text{max}\, \{ |F^{\ibar}|^{1/2} \} \; .
\eeqa
Evaluated for our vacua the F-terms at leading order in
$1/\cV_0,\,1/\cV_{OM,0}$ are
\begin{alignat}{3}
F^{\Sbar} &= -e^{\frac{1}{2}(K_{(A)}+K_{(Z)})}
\sqrt{\frac{6\cV_0^3}{d\cV_{OM,0}^3}} \frac{W_{GC}}{C_H} \\
F^{\Tbar} &= e^{\frac{1}{2}(K_{(A)}+K_{(Z)})}
\sqrt{\frac{2\cV_{OM,0}}{3d\cV_0}}
\Big(\frac{\gamma}{C_H}W_{GC}-W_{OM}\Big) \\
F^{\Ibar} &\equiv F^{\Cbar^{\Ibar}} = \cN
e^{\frac{1}{2}(K_{(A)}+K_{(Z)})}
\Big(\frac{1}{6d\cV_0\cV_{OM,0}}\Big)^{1/2}
\bigg(C^{\Ibar}\Big[\frac{(\gamma-2\beta_v)}{C_H}W_{GC}-W_{OM}\Big]
\\
&\qquad\qquad\, + 3H^{\Ibar J}\Lambda_{JKL} C^K C^L\bigg) \; .
\end{alignat}
Here we have discarded contributions of higher powers in $C$ to
the first two F-terms as they are due to (\ref{CApprox})
exponentially suppressed w.r.t.~the terms given here. As is
clearly recognizable from tables \ref{aTable}-\ref{VTable} our
vacua exhibit next to $\cV_0\gg 1,\,\cV_{OM,0}\gg 1$ the
inequality $\cV_0 > \cV_{OM_0}$. This means that
\beqa
F^{\Sbar} > F^{\Tbar} \gg F^{\Ibar} \; ,
\eeqa
the last inequality once more because of (\ref{CApprox}) which
renders $F^{\Ibar}$ exponentially suppressed as compared to the
former two F-terms. We therefore conclude that
\beqa
M_{SUSY} = M | F^{\Sbar} |^{1/2} = M
e^{\frac{1}{4}(K_{(A)}+K_{(Z)})}
\bigg(\frac{6\cV_0^3}{d\cV_{OM,0}^3}\bigg)^{1/4}
\left(\frac{|W_{GC}|}{C_H}\right)^{1/2} \; .
\eeqa

The next important entity characterizing the de Sitter vacua is
the gravitino mass. Its value is given by
\beqa
m_{3/2}^2 = M^2 e^{K/2} |W| \; .
\eeqa
For its evaluation we will have to determine the absolute value of
the complete superpotential. Note first that because of
(\ref{CApprox}) the cubic $C$ superpotential is again
exponentially smaller as compared to the two non-perturbative
superpotentials and will therefore be discarded. To evaluate the
remaining $|W_{OM}+W_{GC}|$ we use (\ref{MixedTerm}) in the axion
sector with even $n$. The gravitino mass then becomes
\beqa
m_{3/2} = M e^{\frac{1}{4}(K_{(A)}+K_{(Z)})}
\bigg(\frac{3}{8d\cV_0\cV_{OM,0}^3}\bigg)^{1/4}
(|W_{OM}|+|W_{GC}|)^{1/2} \; .
\eeqa
Hence the ratio of supersymmetry breaking scale to gravitino mass
is
\beqa
\frac{M_{SUSY}}{m_{3/2}} = \frac{2\cV_0}{C_H^{1/2}}
\left(\frac{|W_{GC}|}{|W_{OM}|+|W_{GC}|}\right)^{1/2}  \; .
\eeqa
As can be seen from the tables in appendix \ref{Tables} the vacuum
energy, supersymmetry breaking scale and gravitino mass can be
brought into the phenomenological regime for rather small unbroken
hidden gauge groups, not too small $d$, small absolute value for
the Pfaffian $|h|$ and CY volumes $\cV_v\ge 300$. In particular it
will be interesting to better understand the Pfaffians $h$.
Progress in this direction was made in \cite{LOPR},\cite{Pfaff}.

\subsection{The Decompactification Limit}

We have stressed before that our effective 4d potential is valid
in the strongly-coupled regime $\cV\gg 1,\cV_{OM}\gg 1$ but looses
validity towards weak-coupling where $\cL\rightarrow 0$ and
therefore $\cV\rightarrow\cV_v,\cV_{OM}\rightarrow 0$. In this
regime some terms which were neglected here are no longer
suppressed and need to be included while on the other hand some
terms which we included here will become negligible. The prime
reason for us to consider the 4d effective potential in the limit
$\cV\gg 1,\cV_{OM}\gg 1$ was to ensure that volumes (in Planck
units) are big enough to trust the field theory framework and
moreover to be able to neglect multiply wrapped open membrane
instantons about which less is known at present.

However, we are certainly able to study the potential in its
decompactification limit in which $\cV_v \rightarrow \infty, \cL
<\cL_{max} \rightarrow \infty$ as this amounts to sending $\cV
\rightarrow \infty, \cV_{OM} \rightarrow \infty$. Here a flat
background geometry is recovered with vanishing vacuum energy. In
accordance with this we will now verify that our potential indeed
fulfills this expectation and declines towards zero in this limit.
First note that since $\cV$ decreases from $\cV(\cL=0)=\cV_v$
towards $\cV(\cL=\cL_{max})=\cV_v/3$ we have in the
decompactification limit $\cV \simeq \cV_v$. Hence the flux
parameter (cf.~(\ref{GFlux3})) vanishes
\beqa
\cG_v \sim \frac{1}{\cV_v^{1/3}} \rightarrow 0 \; ,
\eeqa
which implies that
\beqa
\cL_{max} \sim \cV_v^{1/3} \rightarrow \infty \; .
\eeqa
It is therefore indeed consistent to study the potential in the
decompactification limit at arbitrarily large orbifold lengths
which is not possible for finite $\cV_v$ since it implies a finite
largest $\cL_{max}$ (the situation of figure \ref{OMGCPotential}).
Now
\beqa
\gamma\cV_{OM} \le \cV_{OM} \sim \cL\cV^{1/3} \le
\cL_{max}\cV^{1/3} \sim \cV^{2/3} \; .
\eeqa
Therefore for the gaugino condensation exponent we get
\beqa
\frac{1}{C_H}(-\cV+\gamma\cV_{OM}) \sim -\frac{\cV}{C_H}
+ \cO(\cV^{2/3}) \; ,
\eeqa
which shows that
\beqa
|W_{GC}| \sim e^{-\cV_v/C_H} \rightarrow 0 \; ,
\qquad |W_{OM}| \sim e^{-\cL\cV_v^{1/3}} \rightarrow 0
\eeqa
and therefore up to polynomial factors the potential energy
density $U$ (\ref{OMGCPot4}) is exponentially declining towards
zero in the decompactification limit as expected. The importance
in showing this smooth connection towards the decompactification
with vanishing energy density lies in the fact that it proves our
de Sitter vacua to be {\it metastable}.

\section{Stabilization of the Non-Universal Moduli}

Up to now we have found a stabilization of the model-independent
moduli through the inclusion of non-perturbative open membrane
instantons and gaugino condensation on the hidden boundary. For
this it was essential that the average CY volume $\cV(\cL)$ and
the orbifold length $\cL$ are not independent. Otherwise one would
have obtained a minimum only in the decompactification limit where
$\cV$ and $\cV_{OM}$ become infinite. This non-trivial volume
dependence was caused by a non-vanishing vev for $G_{lmnp}$. A
non-trivial vev for the four-form $G$ is generically required due
to the boundaries which represent magnetic sources for $G$ and
therefore render the Bianchi-identity for $G$ non-trivial.
However, in general one could also have a $G_{lmn11}$ component
compatible with the boundary sources. Namely the general solution
to the Bianchi identity \beqa dG = \delta (x^{11}-x^{11}_i) S_i(y)
\wedge dx^{11} \eeqa is given by \beqa G =
p\,\Theta(x^{11}-x^{11}_i) S_i(y) +
q\,\delta(x^{11}-x^{11}_i)\omega_i(y)\wedge dx^{11} \eeqa where
$p+q=1$, the Chern-Simons three-forms $\omega_i(y)$ satisfy
$d\omega_i(y)=S_i$ and $S_i$ are the possible magnetic four-form
sources. $G$ therefore contains both types of field-strengths.
Whereas the first part ($G_{lmnp}$) exists through (part of) the
bulk, the second part ($G_{lmn11}$) is localized on the boundaries
or on the M5 branes in case that these are present. This component
of the eleven dimensional flux corresponds to the $H$-flux
appearing in the weakly coupled heterotic string. It was shown in
\cite{NK}, and the second reference of \cite{NoGo} that this flux
acts as a torsion on the internal geometry so that the six
dimensional internal manifold $X$ is no longer K\"ahler and
satisfies $d\omega \neq 0$, where $\omega$ is the fundamental
two-form of the internal manifold. As we already mentioned at the
beginning of this paper, this component of the flux is orthogonal
to the component taken into account in this paper and originates
from the second term appearing in the previous Bianchi identity.

Recall that in \cite{BD}, \cite{TT} and \cite{CL} it was argued
that most of the moduli fields of the non-K\"ahler internal
manifold can be stabilized once the $G_{lmn11}$ component of the
flux has a non-trivial vev, even though the question on how the
moduli fields for this class of manifolds look like very
concretely still remains open. Work in this direction is in
progress. Let us recapitulate the main results for the moduli
stabilization mechanism of the weakly coupled heterotic string
proposed in the previous references as the results obtained
therein very nicely supplement the results obtained in this paper.

The supersymmetry constraints that are present for
compactifications of the weakly heterotic string were first
derived in \cite{NK} and the second reference of \cite{NoGo}. One
of these constraints is the torsional equation which relates the
fundamental two-form of the internal manifold to the $H$-flux
\beqa
H = i(\partial-\bar \partial)\omega \; .
\eeqa The expression for the
$H$-flux that is gauge invariant and anomaly free is written in
terms of the Chern Simons three-forms of the non abelian gauge
field $A$ and the spin connection ${\omega}_0$
\beqa
H=dB-{\alpha}'[{\Omega}_3(A)-{\Omega}_3({\omega}_0-{1\over 2}H)]
\; .
\eeqa
This torsion is not closed because we cannot embed the gauge
connection into the torsional spin connection, as in that case the
$H$-flux would be vanishing.  In order to obtain a solution for
the $H$-flux for a specific background geometry one has to solve
the previous equation iteratively in $H$, as the flux appears on
both sides of the previous equation. This has in done in \cite{BD}
for a particular background of the $SO(32)$ heterotic string that
can be obtained from duality chasing a particular model of the
general class of models describing M-theory compactifications with
non-vanishing fluxes \cite{BB}. What becomes clear from the
previous two equations is that under a rescaling of the
fundamental form with an overall factor `t' which represents the
radial modulus and determines the volume of the internal manifold
\beqa
\omega \rightarrow t \omega \; ,
\eeqa
the $H$-flux does not transform in any simple way, so that it is
expected that the radial modulus can be stabilized for this type
of compactifications at {\em tree level}. In fact, it is expected
that all the complex structure moduli and at least some of the
K\"ahler moduli can be stabilized in these compactifications.
Notice that we are using a rather lax wording at this point,
because as explained before, the actual moduli for this type of
compactifications are still under investigation. The
superpotential responsible for the stabilization of these fields
takes the form (cf.~\cite{BD} fourth reference and \cite{CL}
second reference)
\beqa
W=\int (H+id{\omega})\wedge {\Omega} \; .
\eeqa
In the ordinary CY case we know that the fundamental two-form is
closed $d{\omega}=0$, so that we recover the superpotential for
the heterotic string compactified on a CY three fold conjectured
in \cite{BG} and checked in \cite{BCONS}. From this formula of the
superpotential we observe that the dynamics of the weakly coupled
theory can be described in terms of a {\em complex} three-form
\beqa
{\cH}=H+id{\omega}\; ,
\eeqa
which is anomaly free and gauge invariant. From the kinetic term
of this three-form $\int |{\cal H}|^2$ we can build the scalar
potential $V$ for the moduli fields (cf.~also \cite{CL} second
reference). For the simplest example in which we only have a
radial modulus this potential was written down explicitly in the
last two references of \cite{BD}
\beqa V(t)={t^3\over {\alpha}'}^3-{2{\alpha}'f^4\over
t^3}+{7{\alpha}'^2f^6\over t^6}+\dots, \eeqa where $f$ is the flux
density. One can easily see that this potential has a minimum for
\beqa t=1.288({\alpha}'|f|^2)^{1/3},\eeqa which stabilizes the
radial modulus and thus the volume of the internal manifold in
terms of the flux density.

\bigskip
\noindent {\large \bf Acknowledgements}\\[2ex]
We would like to thank F.~Denef, S.~Kachru, J.~Maldacena,
G.~Moore, J.~Pati, E.~Poppitz, Q.~Shafi, G.~Shiu, L.~Susskind,
H.~Verlinde for interesting discussions related to this work. The
work of M.B.~is supported by NSF grant PHY-01-5-23911 and an
Alfred Sloan Fellowship, that of A.K.~by NSF grant PHY-0099544.

\begin{appendix}

\section{K\"ahler-Potential and Derivatives}

Let's consider the following part of the K\"ahler-potential which
neglects the complex structure and bundle moduli contributions
\begin{alignat}{3}
\Ktil := K_{(S)}+K_{(T)}+K_{(C)} &= -\ln \left(
\frac{d}{6}(S+\Sbar)(T+\Tbar )^3 \right)
+ \left(\frac{3}{T+\Tbar}+\frac{2\beta_v}{S+\Sbar}\right)H_{I\Jbar}
C^I\Cbar^{\Jbar} \notag \\
&= -\ln\left( \frac{8}{3}d\cV\cV_{OM}^3 \right) +
\left(\frac{3}{2\cV_{OM}}+\frac{\beta_v}{\cV}\right)
H_{I\Jbar}C^I\Cbar^{\Jbar}
\end{alignat}
It follows that
\begin{equation}
e^{\Ktil} = \frac{3}{8d\cV\cV_{OM}^3} \left( 1+
\left(\frac{3}{2\cV_{OM}}+\frac{\beta_v}{\cV}\right)
H_{I\Jbar}C^I\Cbar^{\Jbar} \right) \; .
\end{equation}

\noindent First derivatives of $\Ktil$ with respect to the moduli
appear in the K\"ahler covariant derivative $D_i W$. They are
given by (notation: $\Ktil_I = \partial \Ktil/\partial C^I, \;
\Ktil_{\Ibar} = \partial \Ktil/\partial {\Cbar}^{\Ibar}$)
\begin{alignat}{3}
  \Ktil_S &= \Ktil_{\Sbar} = -\frac{1}{2\cV}\left(1+\beta_v
  \frac{H_{I\Jbar}C^I\Cbar^{\Jbar}}{\cV}\right) \\
  \Ktil_T &= \Ktil_{\Tbar} = -\frac{3}{2\cV_{OM}}\left(1+
  \frac{H_{I\Jbar}C^I\Cbar^{\Jbar}}{2\cV_{OM}}\right) \\
  \Ktil_I &= \left(\frac{3}{2\cV_{OM}}+\frac{\beta_v}{\cV}\right)
  H_{I\Jbar}\Cbar^{\Jbar}
\end{alignat}

\noindent Second derivatives:
\begin{alignat}{3}
  \Ktil_{S\Sbar} &= \frac{1}{4\cV^2}
  \left(1+2\beta_v \frac{H_{I\Jbar}C^I\Cbar^{\Jbar}}{\cV}\right)
  \\
  \Ktil_{S\Tbar} &= 0
  \\
  \Ktil_{S\Jbar} &= -\frac{\beta_v H_{I\Jbar}C^I}{2\cV^2}
  \\
  \Ktil_{T\Tbar} &= \frac{3}{4\cV_{OM}^2}
  \left(1+\frac{H_{I\Jbar}C^I\Cbar^{\Jbar}}{\cV_{OM}}\right)
  \\
  \Ktil_{T\Jbar} &= -\frac{3H_{I\Jbar}C^I}{4\cV_{OM}^2}
  \\
  \Ktil_{I\Jbar} &= \left(\frac{3}{2\cV_{OM}}+\frac{\beta_v}{\cV}\right)
                    H_{I\Jbar}
\end{alignat}
These expressions contain leading and subleading terms in
$\delta$. The K\"ahler-matrix ${\bf \Ktil}$ of these second
derivatives of $\Ktil$ then reads
\beqa
\label{K matrix}
{\bf \Ktil} = {\bf K}_0+\delta{\bf K}_1 \; .
\eeqa
Its inversion is given by
\beqa
 {\bf \Ktil}^{-1} = (1-\delta {\bf K}_0^{-1}{\bf K}_1)
 {\bf K}^{-1}_0 + \cO(\delta^2)
\eeqa
and reads in components at leading order (we have dropped all
$\delta^2$ contributions or higher)
\begin{alignat}{3}
  \Ktil^{\Sbar S} &= 4\cV^2 \hspace{4cm}\sim \cV^2
  \\
  \Ktil^{\Sbar T} &= \frac{4}{3}\beta_v\cN\cV_{OM}
  H_{I\Jbar}C^I\Cbar^{\Jbar} \hspace{1cm}\sim \cV_{OM}^2 \; \delta
  \\
  \Ktil^{\Sbar I} &= \frac{4}{3}\beta_v\cN\cV_{OM} C^I
  \hspace{2.2cm} \sim \cV_{OM}^{3/2}\delta^{1/2}
  \\
  \Ktil^{\Tbar T} &= \frac{4\cV_{OM}^2}{3} \hspace{3.5cm}\sim \cV_{OM}^2
  \\
  \Ktil^{\Tbar I} &= \frac{2}{3}\cN\cV_{OM} C^I
  \hspace{2.6cm} \sim \cV_{OM}^{3/2}\delta^{1/2}
  \\
  \Ktil^{\Ibar J} &= \frac{2}{3}\cN\cV_{OM}H^{\Ibar J}
  \hspace{2.35cm} \sim \cV_{OM}
\end{alignat}
where we have defined
\beqa
\cN = \frac{1}{1+\beta_v\frac{2\cV_{OM}}{3\cV}} \; .
\eeqa

\section{Dependence of Minimum's Position on Parameters}
\label{Tables}

We present in the following tables the influence of varying the
hidden gauge group and $d,|h|,\cV_v$ on our de Sitter vacua. The
visible boundary CY volume $\cV_v$ has not been fixed explicitly
in this paper, so we keep it as a parameter, though its
stabilization might be achieved along the lines outlined in
chapter five. More specifically we give the position of the
critical point $\cL_0$, its approximation $\cL_{prox}$ (see
(\ref{SimpleLMin})), the maximal length $\cL_{max}$ for
comparison. Beyond these the values $\cV_0 =
\cV(\cL_0),\,\cV_{OM,0} = \cV_{OM}(\cL_0)$ will be important as
one has to check that they are substantially bigger than 1 to have
our vacua in a controllable regime. Moreover, of clear
phenomenological interest will be the value for the positive
vacuum energy $E$, the supersymmetry breaking scale $\Mtil_{SUSY}$
and the mass of the gravitino $\mtil_{3/2}$ defined as
\beqa
E = ( U/e^{K_{(A)}+K_{(Z)}} )^{1/4} \; , \quad \Mtil_{SUSY}
= M_{SUSY}/e^{K_{(A)}+K_{(Z)}} \; , \quad
\mtil_{3/2} = m_{3/2}/e^{K_{(A)}+K_{(Z)}}
\eeqa
at the critical point. Notice that these values bear also a factor
$e^{-K_{(Z)}}$ (the $K_{(A)}$ being negligible) as compared to the
actual physical parameters $U^{1/4},M_{SUSY},m_{3/2}$. This might
therefore change the values $U^{1/4},M_{SUSY},m_{3/2}$ still by a
few orders of magnitude. In all of the following tables we will
keep the discrete parameter
\beqa
\beta_v = 1
\eeqa
\begin{table}[th]
\begin{center}
\begin{tabular}{|c||c|c|c|c|c|c|c|}
\hline $H,C_H$ & $E_8,30$ & $E_6,12$ & $SO(10),8$ & $SU(5),5$ &
$SU(3),3$ & $SU(2),2$ \\
\hline
\hline $\cL_0$ & $1.1$ & $3.0$ & $4.1$ & $5.5$ & $7.1$ & $8.3$ \\
\hline $\cL_{prox}$ & $1.8$ & $3.6$ & $4.6$ & $6.0$ & $7.5$ & $8.6$ \\
\hline $\cL_{max}$ & $7.2$ & $7.2$ & $7.2$ & $7.2$ & $7.2$ & $7.2$ \\
\hline $\cV_0$ & $341$ & $256$ & $215$ & $171$ & $135$ & $116$ \\
\hline $\cV_{OM,0}$ & $7$ & $16$ & $21$ & $26$ & $31$ & $34$ \\
\hline $E/TeV$ & $1.2\times 10^9$ & $5.7\times 10^6$ & $495213$
& $36455$ & $3185$ & $564$ \\
\hline $\Mtil_{SUSY}/TeV$ & $3.1\times 10^{10}$ & $1.1\times 10^8$
& $8.5\times 10^6$ & $527697$ & $39158$ & $6406$ \\
\hline $\mtil_{3/2}/TeV$ & $3.4\times 10^8$ & $1.3\times 10^6$
& $103752$ & $6991$ & $567$ & $96$ \\
\hline
\end{tabular}
\caption{The dependence on the hidden gauge group $H$ which enters
through its dual Coxeter number $C_H$. The other parameters are
set to $\cV_v=400,d=10,|h|=10^{-8}$.}
\label{aTable}
\end{center}
\end{table}

\noindent fixed at this value. Table \ref{aTable} shows the
dependence on the hidden gauge group. As the dual Coxeter number
$C_H$ of $H$ enters through an exponential, changing $C_H$ can
easily bring $\cL_0$ close to $\cL_{max}$, the phenomenologically
desirable value (see \cite{WWarp},\cite{CK3}). Hence hidden gauge
groups with lower rank are favored. Notice that a hidden $SU(2)$
which is presented in the last column already pushes $\cL_0$
beyond $\cL_{max}$ and would therefore be ruled out for those
parameters chosen.
\begin{table}[!bh]
\begin{center}
\begin{tabular}{|c||c|c|c|c|c|c|c|}
\hline $d$ & 1 & 10 & 50 & 100 & 1000 & 10000 \\ \hline
\hline $\cL_0$ & $3.3$ & $7.1$ & $12.1$ & $15.3$ & $33.0$ & $71.0$ \\
\hline $\cL_{prox}$ & $3.5$ & $7.5$ & $12.8$ & $16.1$ & $34.7$
& $74.7$ \\
\hline $\cL_{max} $ & $3.3$ & $7.2$ & $12.3$ & $15.5$ & $33.3$
& $71.8$ \\
\hline $\cV_0$ & $135$ & $135$ & $135$ & $135$ & $135$ & $135$ \\
\hline $\cV_{OM,0}$ & $31$ & $31$ & $31$ & $31$ & $31$ & $31$ \\
\hline $E/TeV$ & $5663$ & $3185$ & $2130$ & $1791$ & $1007$ & $566$ \\
\hline $\Mtil_{SUSY}/TeV$ & $69634$ & $39158$ & $26187$ & $22020$
& $12383$ & $6963$ \\
\hline $\mtil_{3/2}/TeV$ & $1007$ & $567$ & $379$ & $319$
& $179$ & $101$ \\
\hline
\end{tabular}
\caption{The dependence on the intersection number $d$. The other
parameters are set to $H=SU(3),\cV_v=400, |h|=10^{-8}$.}
\label{dTable}
\end{center}
\end{table}
Table \ref{dTable} illustrates the dependence on the CY
intersection number $d$.
\begin{table}[th]
\begin{center}
\begin{tabular}{|c||c|c|c|c|c|}
\hline $|h|$ & $10^{-14}$ & $10^{-12}$ & $10^{-10}$ & $10^{-8}$ &
$10^{-6}$ \\ \hline
\hline $\cL_0$ & $12.9$ & $13.6$ & $14.4$ & $15.3$ & $16.2$  \\
\hline $\cL_{prox}$ & $16.1$ & $16.1$ & $16.1$ & $16.1$ & $16.1$  \\
\hline $\cL_{max} $ & $15.5$ & $15.5$ & $15.5$ & $15.5$ & $15.5$  \\
\hline $\cV_0$ & $160$ & $151$ & $143$ & $135$ & $127$  \\
\hline $\cV_{OM,0}$ & $27$ & $28$ & $30$ & $31$ & $32$  \\
\hline $E/TeV$ & $10$ & $56$ & $318$ & $1791$ & $9852$  \\
\hline $\Mtil_{SUSY}/TeV$ & $125$ & $706$ & $3969$ & $22020$ & $119823$ \\
\hline $\mtil_{3/2}/TeV$ & $2$ & $10$ & $57$ & $319$ & $1744$ \\
\hline
\end{tabular}
\caption{The dependence on the modulus $|h|$ of the open membrane
Pfaffian. The remaining parameters are set to $H=SU(3), d=100,
\cV_v = 400$.}
\label{hTable}
\end{center}
\end{table}

\noindent Since $\cV_0,\cV_{OM,0}$ are independent of $d$, it
follows from (\ref{VOM}) that
\beqa
\cL_0 = 3.075\, d^{1/3} \; .
\eeqa
With this relation one reproduces the results for $\cL_0$ in table
\ref{dTable}. One sees that larger values for $d$ bring
$E,\Mtil_{SUSY},\mtil_{3/2}$ closer towards the phenomenologically
relevant regime. As to the dependence on the modulus $|h|$ of the
Paffian we see from table \ref{hTable} that clearly small values
are favored. Notice again, that such small values are not too
surprising in view of the bound (\ref{hBound}) on $|h|$. Finally
table \ref{VTable} shows that too small values for $\cV_v$ are
phenomenologically disfavored
\begin{table}[!bh]
\begin{center}
\begin{tabular}{|c||c|c|c|c|c|c|}
\hline $\cV_v$ & $50$ & $100$ & $200$ & $300$ & $400$ & $500$
\\ \hline
\hline $\cL_0$ & $2.6$ & $3.9$ & $5.4$ & $6.4$ & $7.1$ & $7.7$ \\
\hline $\cL_{prox}$ & $3.3$ & $4.5$ & $5.8$ & $6.8$ & $7.5$ & $8.1$ \\
\hline $\cL_{max} $ & $7.2$ & $7.2$ & $7.2$ & $7.2$ & $7.2$ &
$7.2$ \\
\hline $\cV_0$ & $34$ & $56$ & $88$ & $113$ & $135$ & $155$ \\
\hline $\cV_{OM,0}$ & $7$ & $13$ & $20$ & $26$ & $31$ & $35$ \\
\hline $E/TeV$ & $1\times 10^9$ & $5\times 10^7$ & $789773$
& $38817$ & $3185$ & $348$ \\
\hline $\Mtil_{SUSY}/TeV$ & $8\times 10^9$ & $4\times 10^8$
& $8\times 10^6$ & $440333$ & $39158$ & $4600$ \\
\hline $\mtil_{3/2}/TeV$ & $4\times 10^8$ & $1\times 10^7$
& $171633$ & $7495$ & $567$ & $58$ \\
\hline
\end{tabular}
\caption{The dependence on the visible CY volume $\cV_v$. The
remaining parameters are $H=SU(3), d=10, |h|=10^{-8}$.}
\label{VTable}
\end{center}
\end{table}

\noindent (notice however the implicit further factor
$e^{-K_{(Z)}}$ which we haven't evaluated explicitly and which
could help to lower e.g.~the supersymmetry breaking scale further)
while values $\cV_v\gtrsim 400$ bring the supersymmetry breaking
scale and gravitino mass closer to the TeV regime. However a
$\cV_v=500$ would already be too large and bring $\cL_0$ beyond
its upper bound $\cL_{max}$. Moreover, we see that $\cL_{prox}$ in
all these cases gives a rather reliable estimate of the true
position $\cL_0$. This is remarkable in view of the drastic
simplification which reduces (\ref{LMinOMGC}) to
(\ref{SimpleLMin}).

\end{appendix}

 \newcommand{\zpc}[3]{{\sl Z. Phys.} {\bf C\,#1} (#2) #3}
 \newcommand{\npb}[3]{{\sl Nucl. Phys.} {\bf B\,#1} (#2) #3}
 \newcommand{\plb}[3]{{\sl Phys. Lett.} {\bf B\,#1} (#2) #3}
 \newcommand{\prd}[3]{{\sl Phys. Rev.} {\bf D\,#1} (#2) #3}
 \newcommand{\prb}[3]{{\sl Phys. Rev.} {\bf B\,#1} (#2) #3}
 \newcommand{\pr}[3]{{\sl Phys. Rev.} {\bf #1} (#2) #3}
 \newcommand{\prl}[3]{{\sl Phys. Rev. Lett.} {\bf #1} (#2) #3}
 \newcommand{\jhep}[3]{{\sl JHEP} {\bf #1} (#2) #3}
 \newcommand{\jcap}[3]{{\sl JCAP} {\bf #1} (#2) #3}
 \newcommand{\cqg}[3]{{\sl Class. Quant. Grav.} {\bf #1} (#2) #3}
 \newcommand{\prep}[3]{{\sl Phys. Rep.} {\bf #1} (#2) #3}
 \newcommand{\fp}[3]{{\sl Fortschr. Phys.} {\bf #1} (#2) #3}
 \newcommand{\nc}[3]{{\sl Nuovo Cimento} {\bf #1} (#2) #3}
 \newcommand{\nca}[3]{{\sl Nuovo Cimento} {\bf A\,#1} (#2) #3}
 \newcommand{\lnc}[3]{{\sl Lett. Nuovo Cimento} {\bf #1} (#2) #3}
 \newcommand{\ijmpa}[3]{{\sl Int. J. Mod. Phys.} {\bf A\,#1} (#2) #3}
 \newcommand{\rmp}[3]{{\sl Rev. Mod. Phys.} {\bf #1} (#2) #3}
 \newcommand{\ptp}[3]{{\sl Prog. Theor. Phys.} {\bf #1} (#2) #3}
 \newcommand{\sjnp}[3]{{\sl Sov. J. Nucl. Phys.} {\bf #1} (#2) #3}
 \newcommand{\sjpn}[3]{{\sl Sov. J. Particles \& Nuclei} {\bf #1} (#2) #3}
 \newcommand{\splir}[3]{{\sl Sov. Phys. Leb. Inst. Rep.} {\bf #1} (#2) #3}
 \newcommand{\tmf}[3]{{\sl Teor. Mat. Fiz.} {\bf #1} (#2) #3}
 \newcommand{\jcp}[3]{{\sl J. Comp. Phys.} {\bf #1} (#2) #3}
 \newcommand{\cpc}[3]{{\sl Comp. Phys. Commun.} {\bf #1} (#2) #3}
 \newcommand{\mpla}[3]{{\sl Mod. Phys. Lett.} {\bf A\,#1} (#2) #3}
 \newcommand{\cmp}[3]{{\sl Comm. Math. Phys.} {\bf #1} (#2) #3}
 \newcommand{\jmp}[3]{{\sl J. Math. Phys.} {\bf #1} (#2) #3}
 \newcommand{\pa}[3]{{\sl Physica} {\bf A\,#1} (#2) #3}
 \newcommand{\nim}[3]{{\sl Nucl. Instr. Meth.} {\bf #1} (#2) #3}
 \newcommand{\el}[3]{{\sl Europhysics Letters} {\bf #1} (#2) #3}
 \newcommand{\aop}[3]{{\sl Ann. of Phys.} {\bf #1} (#2) #3}
 \newcommand{\jetp}[3]{{\sl JETP} {\bf #1} (#2) #3}
 \newcommand{\jetpl}[3]{{\sl JETP Lett.} {\bf #1} (#2) #3}
 \newcommand{\acpp}[3]{{\sl Acta Physica Polonica} {\bf #1} (#2) #3}
 \newcommand{\sci}[3]{{\sl Science} {\bf #1} (#2) #3}
 \newcommand{\vj}[4]{{\sl #1~}{\bf #2} (#3) #4}
 \newcommand{\ej}[3]{{\bf #1} (#2) #3}
 \newcommand{\vjs}[2]{{\sl #1~}{\bf #2}}
 \newcommand{\hepph}[1]{{\sl hep--ph/}{#1}}
 \newcommand{\desy}[1]{{\sl DESY-Report~}{#1}}

\bibliographystyle{plain}

\begin{thebibliography}{99}

\bibitem{WMAP} D.N.~Spergel et al.,
{\it First Year Wilkinson Microwave Anisotropy Probe (WMAP)
Observations: Determination of Cosmological Parameters},
astro-ph/0302209

\bibitem{GZ} S.~J.~Gates and B.~Zwiebach,
{\it Gauged N=4 Supergravity Theory with a New Scalar Potential},
\plb{123}{1983}{200}; S.~Weinberg, {\it The Cosmological Constant
Problem}, \rmp{61}{1989}{1}

\bibitem{FTP} P.~Fr\'e, M.~Trigiante and A.~Van Proeyen,
{\it N=2 Supergravity Models with Stable De Sitter Vacua},
\cqg{20}{2003}{487}, hep-th/0301024

\bibitem{BS} C.M.~Hull, {\it Domain Wall and De Sitter Solutions
of Gauged Supergravity}, \jhep{0111}{2001}{061}, hep-th/0110048;
K.~Behrndt and S.~Mahapatra, {\it De Sitter Vacua from N=2 Gauged
Supergravity}, hep-th/0312063

\bibitem{GdS} S.J.~Gates, {\it Is Stringy Supersymmetry
Quintessentially Challenged?}, hep-th/0202112

\bibitem{KKLT} S.~Kachru, R.~Kallosh, A.~Linde and S.P.~Trivedi,
{\it De Sitter Vacua in String Theory}, \prd{68}{2003}{046005},
hep-th/0301240

\bibitem{dSLit} V.~Balasubramanian, P.~Horava and D.~Minic, {\it
Deconstructing De Sitter}, \jhep{0105}{2001}{043}, hep-th/0103171;
C.M.~Hull, {\it De Sitter Space in Supergravity and M-Theory},
\jhep{0111}{2001}{012}, hep-th/0109213; G.W.~Gibbons and
C.M.~Hull, {\it De Sitter Space From Warped Supergravity
Solutions}, hep-th/0111072; P.~Berglund, T.~H\"ubsch and D.~Minic,
{\it De Sitter Space-Times From Warped Compactifications of IIB
String-Theory}, \plb{534}{2002}{147}, hep-th/0112079; M.~Fabinger
and E.~Silverstein, {\it D-Sitter Space: Causal Structure,
Thermodynamics, And Entropy}, hep-th/0304220; E.~Silverstein, {\it
AdS and dS Entropy From String Junctions: Or the Function of
Junction Conjunctions}, hep-th/0308175; C.~Escoda, M.~Gomez-Reino
and F.~Quevedo, {\it Saltatory De Sitter String Vacua},
\jhep{0311}{2003}{065}, hep-th/0307160; H.~Lu and
J.F.~Vazquez-Poritz, {\it Four-Dimensional Einstein-Yang-Mills-De
Sitter Gravity From Eleven Dimensions}, hep-th/0308104;
C.P.~Burgess, R.~Kallosh and F.~Quevedo, {\it De Sitter Vacua from
Supersymmetric D-Terms}, \jhep{0310}{2003}{056}, hep-th/0309187;
R.~Brustein and S.P.~de~Alwis, {\it Moduli Potentials in String
Compactifications with Fluxes: Mapping the Discretuum},
hep-th/0402088

\bibitem{SS} A.~Saltman and E.~Silverstein,
{\it The Scaling of the No-Scale Potential and de Sitter Model
Building}, hep-th/0402135

\bibitem{NK}
A.~Strominger, {\it Superstrings with Torsion},
\npb{274}{1986}{253}

\bibitem{NoGo} G.W.~Gibbons,
{\it Aspects of Supergravity Theories}, published in {\it GIFT
Seminar on Supersymmetry, Supergravity and Related Topics},
eds.~F.~del Aguila, J.A.~de Azcarraga and L.E.~Ibanez, World
Scientific (1984); B.~de Wit, D.J.~Smit and N.D.~Hari Dass, {\it
Residual Supersymmetry of Compactified $D=10$ Supergravity},
\npb{283}{1987}{165}; J.M.~Maldacena and C.~Nunez, {\it
Supergravity Description of Field Theories on Curved Manifolds and
a No-Go Theorem}, \ijmpa{16}{2001}{822}, hep-th/0007018; N.D.~Hari
Dass, {\it A No Go Theorem for De Sitter Compactifications?},
\mpla{17}{2002}{1001}, hep-th/0205056;

\bibitem{CK2} G. Curio and A. Krause,
{\it G-Fluxes and Non-Perturbative Stabilization of Heterotic
M-Theory}, \npb{643}{2002}{131}, hep-th/0108220

\bibitem{TW} P.K.~Townsend and M.N.R.~Wohlfarth,
{\it Accelerating Cosmologies from Compactification},
\prl{91}{2003}{061302}, hep-th/0303097; M.N.~Wohlfarth,
{Accelerating Cosmologies and a Phase Transition in M-Theory},
\plb{563}{2003}{1}, hep-th/0304089

\bibitem{GS} M.~Gutperle and A.~Strominger, {\it Space-Like
Branes}, \jhep{0204}{2002}{018}, hep-th/0202210

\bibitem{NO} N.~Ohta, {\it Accelerating Cosmologies From S-Branes},
\prl{91}{2003}{061303}, hep-th/0303238

\bibitem{SBC} S.~Roy, {\it Accelerating Cosmologies from
M/String-Theory Compactifications}, \plb{567}{2003}{322},
hep-th/0304084; R.~Emparan and J.~Garriga, {\it A Note on
Accelerating Cosmologies from Compactifications and S-Branes},
\jhep{0305}{2003}{028}, hep-th/0304124; N.~Ohta, {\it A Study of
Accelerating Cosmologies from Superstring/M-Theories},
\ptp{110}{2003}{269}, hep-th/0304172; M.~Gutperle, R.~Kallosh and
A.~Linde, {\it M/String-Theory, S-Branes and Accelerating
Universe}, \jcap{0307}{2003}{001}, hep-th/0304225; C.M.~Chen,
P.M.~Ho, I.P.~Neupane, N.~Ohta and J.E.~Wang, {\it Hyperbolic
Space Cosmologies}, \jhep{0310}{2003}{058}, hep-th/0306291

\bibitem{DHHK} K.~Dasgupta, C.~Herdeiro, S.~Hirano and R.~Kallosh,
{\it D3/D7 Inflationary Model and M Theory}, hep-th/0203019

\bibitem{GKP} S.B.~Giddings, S.~Kachru and J.~Polchinski, {\it
Hierarchies from Fluxes in String Compactifications},
\prd{66}{106006}{2002}, hep-th/0105097

\bibitem{WWarp} E.~Witten,
{\it Strong Coupling Expansion of Calabi-Yau Compactification},
\npb{471}{1996}{135}, hep-th/9602070

\bibitem{HetMPheno} H.P.~Nilles, M.~Olechowski and M.~Yamaguchi,
{\it Supersymmetry Breaking and Soft Terms in M-Theory},
\plb{415}{1997}{24}, hep-th/9707143; H.P.~Nilles, M.~Olechowski
and M.~Yamaguchi, {\it Supersymmetry Breakdown at a Hidden Wall},
\npb{530}{1998}{43}, hep-th/9801030; R.~Arnowitt and B.~Dutta,
{\it Yukawa Textures and Horava-Witten M-Theory},
\ijmpa{16S1C}{2001}{940}, hep-th/0008238; A.E.~Faraggi and
R.S.~Garavuso, {\it Yukawa Couplings in SO(10) Heterotic M-Theory
Vacua}, \npb{659}{2003}{224}, hep-th/0301147; R.~Arnowitt,
B.~Dutta and B.~Hu, {\it Quark, Lepton and Neutrino Masses in
Horava-Witten Inspired Models}, hep-ph/0211084; {\it Yukawa
Textures, Neutrino Masses and Horava-Witten M-Theory},
hep-th/0309033;

\bibitem{CK1} G.~Curio and A.~Krause,
{\it Four-Flux and Warped Heterotic M-Theory Compactifications},
\npb{602}{2001}{172}, hep-th/0012152

\bibitem{CK3} G.~Curio and A.~Krause,
{\it Enlarging the Parameter Space of Heterotic M-Theory Flux
Compactifications to Phenomenological Viability}, {\sl Nucl.
Phys.} {\bf B} in print, hep-th/0308202

\bibitem{BD} K.~Dasgupta and S.~Sethi, {\it M Theory, Orientifolds
and G-Flux}, \jhep{9908}{1999}{023}, hep-th/9908088; K.~Becker and
K.~Dasgupta, {\it Heterotic Strings with Torsion},
\jhep{0211}{2002}{006}, hep-th/0209077; K.~Becker, M.~Becker,
K.~Dasgupta and P.~S.~Green, {\it Compactifications of Heterotic
String Theory on Non-K\"ahler Complex Manifolds.I},
\jhep{0304}{2003}{0304}, hep-th/0301161; K.~Becker, M.~Becker,
K.~Dasgupta and S.~Prokushkin, {\it Properties of Heterotic Vacua
from Superpotentials}, \npb{666}{2003}{144}, hep-th/0304001;
K.~Becker, M.~Becker, K.~Dasgupta, P.~S.~Green and E.~Sharpe, {\it
Compactifications of Heterotic String Theory on Non-K\"ahler
Complex Manifolds.II}, \npb{678}{2004}{19}, hep-th/0310058

\bibitem{TT} P.~K.~Tripathy and S.~P.~Trivedi, {\it Compactification
with Flux on K3 and Tori}, \jhep{0303}{2003}{028}, hep-th/0301139

\bibitem{CL} G.~L.~Cardoso, G.~Curio, G.~Dall'Agata, D.~L\"ust,
P.~Manousselis and G.~Zoupanos, {\it Non-K\"ahler String
Backgrounds and their Five Torsion Classes}, \npb{652}{2003}{5},
hep-th/0211118; G.~L.~Cardoso, G.~Curio, G.~Dall'Agata and
D.~L\"ust, {\it BPS Action and Superpotential for Heterotic String
Compactifications with Fluxes}, \jhep{0310}{2003}{004},
hep-th/0306088; {\it Heterotic String Theory on Non-K\"ahler
Manifolds with H-Flux and Gaugino Condensate}, hep-th/0310021

\bibitem{GKLA} S.~Gukov, S.~Kachru, X.~Liu and L.~McAllister, {\it
Heterotic Moduli Stabilization with Fractional Chern-Simons
Invariants }, hep-th/0310159

\bibitem{dS1} N.~Goheer, M.~Kleban and L.~Susskind, {\it The
Trouble with de Sitter Space}, \jhep{0307}{2003}{056},
hep-th/0212209;

\bibitem{dS2} S.B.~Giddings, {\it The Fate of Four Dimensions},
hep-th/0303031

\bibitem{dS3} T.~Banks, {\it Some Thoughts on the Quantum Theory of de Sitter
Space}, astro-ph/0305037

\bibitem{KSUSY03} A.~Krause, {\it On de Sitter Vacua in Strongly
Coupled Heterotic String Theory}, talk at SUSY '03, hep-th/0404001

\bibitem{BO} E.~I.~Buchbinder and B.~A.~Ovrut, {\it Vacuum
Stability in Heterotic M-Theory}, hep-th/0310112

\bibitem{Nano} J.~Ellis, C.~Kounnas and D.~V.~Nanopoulos,
{\it No-Scale Supergravity GUTs}, \npb{247}{1984}{373}

\bibitem{BinGai} P.~Binetruy and M.~K.~Gaillard,
{\it Radiative Corrections in Compactified Superstring Models},
\plb{168}{1986}{347}

\bibitem{HW1} P.~Ho\v{r}ava and E.~Witten,
{\it Heterotic and Type I String Dynamics from Eleven Dimensions},
\npb{460}{1996}{506}, hep-th/9510209

\bibitem{HW2} P.~Ho\v{r}ava and E.~Witten,
{\it Eleven-Dimensional Supergravity on a Manifold with Boundary},
\npb{475}{1996}{94}, hep-th/9603142

\bibitem{LOW1} A.~Lukas, B.A.~Ovrut and D.~Waldram,
{\it On the Four-Dimensional Effective Action of Strongly Coupled
Heterotic String Theory}, \npb{532}{1998}{43}, hep-th/9710208

\bibitem{MPS} G.~Moore, G.~Peradze and N.~Saulina,
{\it Instabilities in Heterotic M-Theory induced by Open Membrane
Instantons}, \npb{607}{2001}{117}, hep-th/0012104

\bibitem{LOW4} A.~Lukas, B.A.~Ovrut and D.~Waldram,
{\it Five-Branes and Supersymmetry Breaking in M-Theory},
\jhep{9904}{1999}{009}, hep-th/9901017

\bibitem{K1} A.~Krause, {\it Heterotic M-Theory, Warped Geometry and
the Cosmological Constant Problem}, \fp{49}{2001}{163};

\bibitem{BD1} T.~Banks and M.~Dine,
{\it Couplings and Scales in Strongly Coupled Heterotic String
Theory}, \npb{479}{1996}{173}, hep-th/9605136

\bibitem{ConstK} D.~Constantin and A.~Krause, {\it work in
progress}

\bibitem{LOSW} A.~Lukas, B.~A.~Ovrut, K.~S.~Stelle and D.~Waldram,
{\it Heterotic M-Theory in Five Dimensions}, \npb{552}{1999}{246},
hep-th/9806051

\bibitem{LOPR} E.~Lima, B.A.~Ovrut, J.~Park and R.~Reinbacher,
{\it Nonperturbative Superpotential from Membrane Instantons in
Heterotic M-Theory}, \npb{614}{2001}{117}, hep-th/0101049

\bibitem{GC1} S.~Ferrara, L.~Girardello and H.P.~Nilles,
{\it Breakdown of Local Supersymmetry through Gauge Fermion
Condensates}, \plb{125}{1983}{457}; J.P.~Derendinger,
L.E.~Ib\'a\~nez and H.P.~Nilles, {\it On the Low-Energy D=4, N=1
Supergravity Theory Extracted from the D=10, N=1 Superstring},
\plb{155}{1985}{65}; {\it On the Low-Energy Limit of Superstring
Theories}, \npb{267}{1986}{365}; P.~Horava, {\it Gluino
Condensation in Strongly Coupled Heterotic String Theory},
\prd{54}{1996}{7561}, hep-th/9608019;

\bibitem{GC2} M.~Dine, R.~Rohm, N.~Seiberg and E.~Witten, {\it Gluino
Condensation in Superstring Models}, \plb{156}{1985}{55};
A.~Lukas, B.A.~Ovrut and D.~Waldram, {\it Gaugino Condensation in
M-Theory on $S^1/Z_2$}, \prd{57}{1998}{7529}, hep-th/9711197;

\bibitem{BD2} T.~Banks and M.~Dine, {\it Phenomenology of Strongly
Coupled Heterotic String Theory}, hep-th/9609046

\bibitem{KPot} A.~Krause, {\it Testing Stability of M-Theory on an
$S^1/Z_2$ Orbifold}, \jhep{0005}{2000}{046}, hep-th/9909182

\bibitem{RM} R.N.~Mohapatra, {\it Unification and Supersymmetry},
Springer-Verlag (2003)

\bibitem{GCRev} H.P.~Nilles, {\it Gaugino Condensation and SUSY
Breakdown}, hep-th/0402022

\bibitem{K3} A.~Krause, {\it A Small Cosmological Constant, Grand
Unification and Warped Geometry}, hep-th/0006226; {\it A Small
Cosmological Constant and Back-Reaction of Nonfinetuned
Parameters}, hep-th/0007233, \jhep{0309}{2003}{016}.

\bibitem{PolBook} J.~Polchinski, {\it String-Theory, Vol.2},
Cambridge University Press (1998);

\bibitem{BHE} A.~Krause, {\it Dual Brane Pairs and the
Bekenstein-Hawking Entropy of Non-Susy Spacetimes},
\cqg{20}{2003}{S533}; {\it On the Bekenstein-Hawking Entropy,
Non-Commutative Branes and Logarithmic Corrections},
hep-th/0312309; {\it Black Holes, Space-Filling Chains and Random
Walks}, hep-th/0312311

\bibitem{Pfaff} E.~Buchbinder, R.~Donagi and B.A.~Ovrut, {\it
Vector Bundle Moduli and Small Instanton Transitions},
\jhep{0206}{2002}{054}, hep-th/0202084; {\it Superpotentials For
Vector Bundle Moduli}, \npb{653}{2003}{400}, hep-th/0205190; {\it
Vector Bundle Moduli Superpotentials in Heterotic Superstrings and
M-Theory}, \jhep{0207}{2002}{066}, hep-th/0206203

\bibitem{BB} K.~Becker and M.~Becker , {\it M-Theory on Eight Manifolds},
\npb{477}{1996}{155}, hep-th/9605053;

\bibitem{BG} K.~Behrndt and S.~Gukov,
{\it Domain Walls and Superpotentials from M-Theory on Calabi-Yau
Three-Folds}, \npb{580}{2000}{225}, hep-th/0001082

\bibitem{BCONS} M.~Becker and D.~Constantin,
{\it A Note on Flux Induced Superpotential in String Theory},
\jhep{0308}{2003}{015}, hep-th/0210131

\end{thebibliography}

\end{document}